\UseRawInputEncoding 
\documentclass[superscriptaddress,aps,prl,reprint,twocolumn,amsmath,amssymb,showpacs,]{revtex4-2}
\usepackage{graphicx}  
\usepackage{mathptmx}
\usepackage{epstopdf}
\usepackage{dcolumn}   
\usepackage{bm}        
\usepackage{amssymb}   
\usepackage{amsmath}   
\usepackage{amsthm}
\usepackage{chemarrow}
\usepackage{color}
\usepackage{mathrsfs}
\usepackage{float}
\usepackage{bbold}
\usepackage{bbm}
\usepackage[a4paper,colorlinks=true,
linkcolor=blue,citecolor=blue,
pdfauthor={ },
pdftitle={ },
pdfsubject={ },
pdfkeywords={ }]{hyperref}

\theoremstyle{plain}
\newtheorem*{theorem*}{Theorem}

\usepackage[usenames,dvipsnames]{xcolor}
\colorlet{BLUE}{blue}
\colorlet{RED}{red}

\begin{document}


\title{Limits on axions and axionlike particles within the axion window using a spin-based amplifier}

\date{\today}

\author{Yuanhong Wang}
\email[]{These authors contributed equally to this work}
\affiliation{
CAS Key Laboratory of Microscale Magnetic Resonance and School of Physical Sciences, University of Science and Technology of China, Hefei, Anhui 230026, China}
\affiliation{
CAS Center for Excellence in Quantum Information and Quantum Physics, University of Science and Technology of China, Hefei, Anhui 230026, China}

\author{Haowen Su}
\email[]{These authors contributed equally to this work}
\affiliation{
CAS Key Laboratory of Microscale Magnetic Resonance and School of Physical Sciences, University of Science and Technology of China, Hefei, Anhui 230026, China}
\affiliation{
CAS Center for Excellence in Quantum Information and Quantum Physics, University of Science and Technology of China, Hefei, Anhui 230026, China}

\author{Min Jiang}
\email[]{dxjm@ustc.edu.cn}
\affiliation{
CAS Key Laboratory of Microscale Magnetic Resonance and School of Physical Sciences, University of Science and Technology of China, Hefei, Anhui 230026, China}
\affiliation{
CAS Center for Excellence in Quantum Information and Quantum Physics, University of Science and Technology of China, Hefei, Anhui 230026, China}

\author{Ying Huan}
\affiliation{
CAS Key Laboratory of Microscale Magnetic Resonance and School of Physical Sciences, University of Science and Technology of China, Hefei, Anhui 230026, China}
\affiliation{
CAS Center for Excellence in Quantum Information and Quantum Physics, University of Science and Technology of China, Hefei, Anhui 230026, China}

\author{\mbox{Yushu Qin}}
\affiliation{
CAS Key Laboratory of Microscale Magnetic Resonance and School of Physical Sciences, University of Science and Technology of China, Hefei, Anhui 230026, China}
\affiliation{
CAS Center for Excellence in Quantum Information and Quantum Physics, University of Science and Technology of China, Hefei, Anhui 230026, China}

\author{\mbox{Chang Guo}}
\affiliation{
CAS Key Laboratory of Microscale Magnetic Resonance and School of Physical Sciences, University of Science and Technology of China, Hefei, Anhui 230026, China}
\affiliation{
CAS Center for Excellence in Quantum Information and Quantum Physics, University of Science and Technology of China, Hefei, Anhui 230026, China}

\author{\mbox{Zehao Wang}}
\affiliation{
CAS Key Laboratory of Microscale Magnetic Resonance and School of Physical Sciences, University of Science and Technology of China, Hefei, Anhui 230026, China}
\affiliation{
CAS Center for Excellence in Quantum Information and Quantum Physics, University of Science and Technology of China, Hefei, Anhui 230026, China}

\author{Dongdong Hu}
\affiliation{
State Key Laboratory of Particle Detection and Electronics, University of Science and Technology of China, Hefei, Anhui 230026, China}

\author{Wei Ji}
\affiliation{Helmholtz-Institut, GSI Helmholtzzentrum f{\"u}r Schwerionenforschung, Mainz 55128, Germany}
\affiliation{Johannes Gutenberg University, Mainz 55128, Germany}

\author{Pavel~Fadeev}
\affiliation{Helmholtz-Institut, GSI Helmholtzzentrum f{\"u}r Schwerionenforschung, Mainz 55128, Germany}
\affiliation{Johannes Gutenberg University, Mainz 55128, Germany}

\author{Xinhua Peng}
\email[]{xhpeng@ustc.edu.cn}
\affiliation{
CAS Key Laboratory of Microscale Magnetic Resonance and School of Physical Sciences, University of Science and Technology of China, Hefei, Anhui 230026, China}
\affiliation{
CAS Center for Excellence in Quantum Information and Quantum Physics, University of Science and Technology of China, Hefei, Anhui 230026, China}

\author{Dmitry Budker}
\affiliation{Helmholtz-Institut, GSI Helmholtzzentrum f{\"u}r Schwerionenforschung, Mainz 55128, Germany}
\affiliation{Johannes Gutenberg University, Mainz 55128, Germany}
\affiliation{Department of Physics, University of California, Berkeley, CA 94720-7300, USA}

\begin{abstract}
Searches for the axion and axionlike particles may hold the key to unlocking some of the deepest puzzles about our universe, such as dark matter and dark energy. 
Here we use the recently demonstrated spin-based amplifier to constrain such hypothetical particles within the well-motivated ``axion window'' (1~$\mu$eV-1~meV) through searching for an exotic spin-spin interaction between polarized electron and neutron spins.
The key ingredient is the use of hyperpolarized long-lived $^{129}$Xe nuclear spins as an amplifier for the pseudomagnetic field generated by the exotic interaction.
Using such a spin sensor,
we obtain a direct upper bound on the product of coupling constants $g_p^e g_p^n$.
The spin-based amplifier technique can be extended to searches for a wide variety of hypothetical particles beyond the Standard Model.
\end{abstract}

\maketitle

\emph{Introduction}.~The possible existence of a dark sector of new particles that could mediate exotic long-range interactions between the visible sector of elementary particles is predicted by numerous theories beyond the Standard Model of particle physics~\cite{demille2017probing,ramsey1979tensor,wilczek1978problem,peccei1977cp,weinberg1978new,fadeev2019revisiting,safronova2018search}. 
These hypothetical particles weakly coupled to the Standard-Model particles have been defined as WISPs (weakly interacting subelectronvolt particles)~\cite{kim2019experimental,yan2013new} including the most plausible mediators, 
axions~\cite{jiang2021search,peccei1977cp,wilczek1978problem} and the dark photon~\cite{jaeckel2010low,langacker2009physics}.
Axions are introduced as a compelling solution to the ``strong CP'' problem of quantum chromodynamics~\cite{peccei1977cp,wilczek1978problem,kim2010axions}.
Axions and axionlike particles (collectively referred to as axions) can generically arise in theories at ultrahigh energies, including string theory, grand unified theories and extra-dimensions models~\cite{irastorza2018new,svrcek2006axions}.
Importantly, axions are prominent dark-matter and dark-energy candidates and thus may indeed explain additional puzzling observations including the rotation curves of galaxies and alignment in the multipoles of the cosmic microwave-background anisotropies~\cite{safronova2018search,kamionkowski2014dark,langacker2009physics,afach2021search}.  
Several theoretical models put forward to explain the experimental and astrophysical observations favor the axion mass in the so-called ``axion window'' (1~$\mu$eV-1~meV)~\cite{turner1990windows,youdin1996limits,arvanitaki2014resonantly}. 

Since the first experimental searches for exotic spin-spin interactions~\cite{ramsey1979tensor}, recent progress on exploring such interactions over a broad range of particle masses has been enabled by advances in techniques including magnetometry~\cite{almasi2020new,ji2018new,kim2019experimental,jiang2020interference,bulatowicz2013laboratory}, nuclear magnetic resonance~\cite{aggarwal2020characterization,arvanitaki2014resonantly,su2021search,jiang2021search}, torsion-pendulum measurements~\cite{heckel2013limits,terrano2015short}, trapped ions measurements~\cite{wineland1991search}, nitrogen-vacancy centers in diamond~\cite{rong2018searching}, geoelectrons~\cite{hunter2013using,hunter2014using}, neutron beams~\cite{yan2013new}, masers~\cite{glenday2008limits,jiang2019floquet} and cantilevers~\cite{ding2020constraints}.
The tree-level interactions arising from pseudoscalar boson
exchange are spin-dependent~\cite{dobrescu2006spin}.
Experiments typically employ a ``spin source'' consisting of a large collection of spin-polarized fermions and a ``spin sensor'' measuring the energy shifts induced by exotic spin-spin interactions.
Following the earlier axion searches  ~\cite{safronova2018search,demille2017probing,kim2010axions,jiang2021search,jaeckel2010low},
we focus on the Yukawa-like exotic spin-spin interaction $V_{pp}$~\cite{fadeev2019revisiting} (equivalently $V_{3}$ in Dobrescu et al.~\cite{dobrescu2006spin}) mediated by axions ($c=\hbar=1$),
\begin{flalign}
      \nonumber
      V_{pp}&=-\frac{g_p^e g_p^n}{4}\left[(\hat{\bm{\sigma}}_1\cdot\hat{\bm{\sigma}}_2)\left(\frac{m_a}{ r^2}+\frac{1}{r^{3}}\right)\right.\notag\\
      &-\left.(\hat{\bm{\sigma}}_1\cdot\hat{r})(\hat{\bm{\sigma}}_2\cdot\hat{r})\left(\frac{m_a^2}{r}+\frac{3m_a}{r^2}+\frac{3}{r^3}\right)\right]\frac{e^{-m_a r}}{4\pi m_{1}m_{2} },
\label{v3}
\end{flalign}
where $g_p^e g_p^n$ is the product of electron and neutron pseudoscalar coupling constants for $V_{pp}$, $\hat{\bm{\sigma}}_i$ is the spin vector of $i$th fermion and $m_i$ is its mass, $r$ is the distance between the two interacting fermions, $\hat{r}$ is the corresponding unit vector, $m_a$ is the axion mass.
The axion window (1~$\mu$eV-1~meV) corresponds to a force range from 0.2~mm to 20~cm.
In this range, the spin source generates a measurable magnetic field on the spin sensor, presenting a challenge of distinguishing the sought-after signal from the effect of this magnetic field.
Former investigations~\cite{wineland1991search,almasi2020new} focused on the relatively long force range and their constraints were thus less stringent within the axion window because the Yukawa-like interaction $V_{pp}$ falls off exponentially with increasing distance.  


In this Letter, we use a spin-based amplifier~\cite{jiang2021search,su2021search} to constrain hypothetical axions within the axion window through searching for the exotic spin-spin interaction $V_{pp}$ between polarized electron and neutron spins.
This spin sensor employs an ensemble of hyperpolarized long-lived $^{129}$Xe nuclear spins as an amplifier for the pseudomagnetic fields from the exotic interactions,
which are read out with an atomic magnetometer~\cite{jiang2020interference,jiang2019magnetic}.
We demonstrate that the signal from pseudomagnetic fields is enhanced by a factor of more than 40. 
Using the spin-based amplifier, we obtain a direct upper bound on $|g_p^e g_p^n|$ for pseudoscalars  and reach into unexplored parameter space for the axion mass from 0.03\,meV to 1\,meV within the axion window.
Although demonstrated for the indirect axion searches, our sensing technique can be extended to resonantly search for hypothetical particles beyond the Standard Model present in the ambient space, such as new spin-1 bosons~\cite{su2021search}, dark photons~\cite{an2015direct} and axionlike particles~\cite{jiang2021search}.

\begin{figure}[t]  
	\makeatletter
	\def\@captype{figure}
	\makeatother
	\includegraphics[scale=1.15]{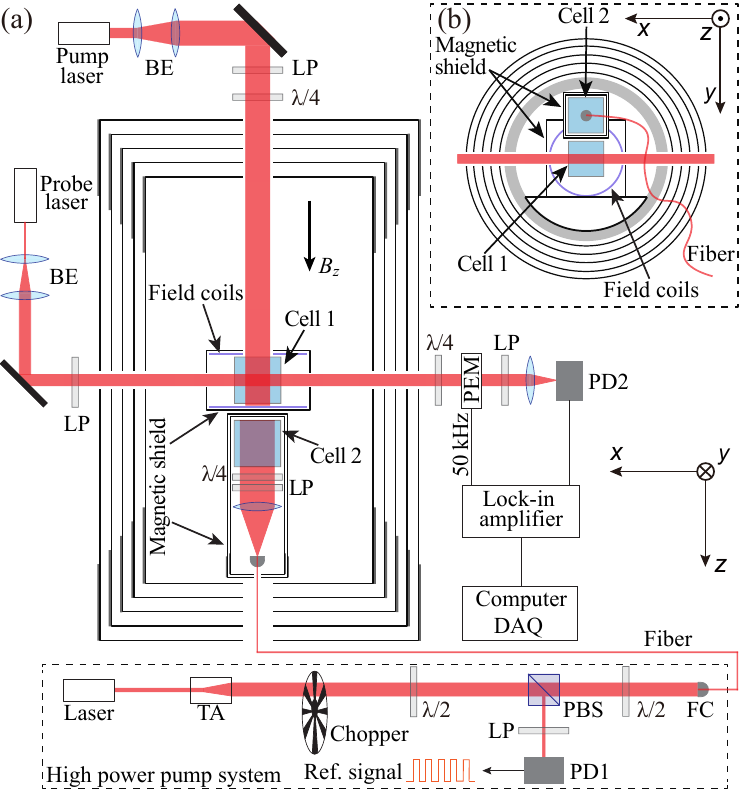}
	\caption{The experimental setup consists of a spin-based amplifier and a spin source. (a) Experimental schematic in the $xz$ plane. The spin source (cell 2) is shielded by a small-size two-layer magnetic shield. The spin sensor (cell 1) is shielded by a small-size one-layer magnetic shield. Both the source and the sensor are enclosed in five-layer magnetic shield. 
	The cell 1~\cite{jiang2020interference,jiang2019floquet,jiang2019magnetic} containing 5~torr of isotopically enriched $^{129}$Xe, 250~torr $\textrm{N}_2$ and a droplet of $^{87}$Rb is heated to 165$^\circ$C. The $^{87}$Rb spins are polarized with a circularly polarized beam of 795\,nm D1 light. $^{129}$Xe spins are polarized to $\sim$30$\%$ in spin-exchange collisions with polarized $^{87}$Rb spins~\cite{walker1997spin,jiang2021search}.
    The $x$ component of $^{87}$Rb spins is measured via optical rotation of a linearly polarized probe beam, which is detuned to higher frequencies by 110\,GHz from the D2 resonance. (b) Experimental schematic in the $xy$ plane. BE, beam expander; LP, linear polarizer; $\lambda /4$, quarter-wave plate; $\lambda /2$, half-wave plate; PD, photodiode; TA, tapered amplifier; PBS, polarizing beam splitter; FC, fiber coupler; PEM, photoelastic modulator; DAQ, data acquisition.}
	\label{setup}
\end{figure}

\emph{Spin source}.~To search for exotic spin-dependent interaction $V_{pp}$, we use a spin source (cell 2) consisting of optically pumped $^{87}$Rb atoms.
The 0.58-cm$^3$ cell 2 contains a droplet of $^{87}$Rb metal and 0.37~amg of N$_2$ gas. The cell is heated to 180$^{\circ}$C.
A 795\,nm pump laser [see Fig.\,\ref{setup}(a)] tuned to the peak of the D1 transition is amplified with a tapered amplifier and then coupled into a single-mode fiber for optical pumping along $\hat{z}$, delivering approximately 0.5\,W of power to the spin-source cell 2.
An optical chopper is used to periodically block the pump beam at a frequency $\nu \approx 10.00$\,Hz with 50$\%$ duty cycle thus modulating the polarization of $^{87}$Rb atoms.
The modulation frequency is monitored with a photodetector.
With pump light on, the polarization of electrons in the whole spin-source cell is averaged over 90$\%$ and the corresponding number of polarized $^{87}$Rb electron spins is $2.18\times 10^{14}$~\cite{SI}. 
The spin-source cell is located 39\,mm away from the center of the spin-based amplifier cell (cell 1), as shown in Fig.\,\ref{setup}(b).

The exotic spin-spin interaction $V_{pp}$ between the source electrons and the spin-sensor $^{129}$Xe neutrons mediated by axions generates a pseudomagnetic field on $^{129}$Xe spins~\cite{ni1999search,su2021search},
\begin{flalign}
     \textbf{B}_{a}&=\frac{g_p^e g_p^n}{16\pi m_{1}m_{2} \mu_{\textrm{Xe}}} \int_{V} \rho(\bm{r}) \left[ \hat{\bm{\sigma}}_2\left( \frac{m_a}{r^2} +\frac{1}{r^{3}}\right)\right.\notag
     \\
     \nonumber
      &\left.-\hat{r}(\hat{\bm{\sigma}}_2\cdot\hat{r})\left(\frac{m_a^2}{r}+\frac{3m_a}{r^2}+\frac{3}{r^3}\right)\right]e^{-m_a r}
      d\bm{r}, \\
      &=\sum \limits_{N} \textbf{B}_{N} \cos(2 \pi N \nu t+\phi_N),
\label{B3}
\end{flalign}
where $\mu_{\textrm{Xe}}$ is the magnetic moment of $^{129}$Xe spin, $\bm{r}$ is the position of the electrons with respect to the sensor, $V$ is the volume of the spin-source cell, $\rho(\bm{r})$ is the spin-density of electrons.
Here, we assume the fractional contribution of neutrons in $^{129}$Xe spins to be 0.73~\cite{kimball2015nuclear}.
Due to the polarization gradient in the spin source, we use a finite element analysis to evaluate the pseudomagnetic field~\cite{SI}.
Because the $^{87}$Rb electron relaxation time ($\sim1$\,ms) is much shorter than the modulation period ($\sim100$\,ms), the modulated electron polarization changes quickly and thus the corresponding pseudomagnetic field can be described by a square wave. 
Based on numerical simulation of Eq.\,\eqref{B3}, we show that the pseudomagnetic field $\textbf{B}_{a}$ can be decomposed into several harmonics i.e., $\nu,2\nu,3\nu,$..., where $\textbf{B}_{N}$ is the $N$th field and $\phi_N$ is its phase~\cite{SI}.

\begin{figure}[t]  
	\makeatletter
	\def\@captype{figure}
	\makeatother
	\includegraphics[scale=1.05]{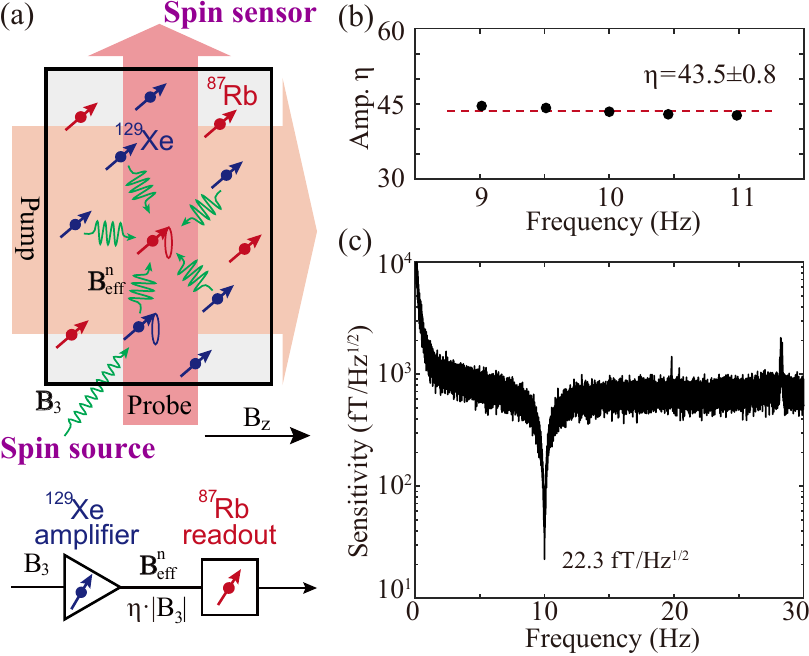}
	\caption{Magnetic-field amplification of the spin sensor. (a) Principle of using the spin sensor to search for exotic interactions. The signal from the pseudomagnetic field is enhanced by the $^{129}$Xe-based amplifier, generating an effective magnetic field $\textbf{B}_{\textrm{eff}}$ read out by $^{87}$Rb spins. (b) The amplification factor $43.5\pm0.8$ is calibrated at frequencies about {9.00, 9.50, 10.00, 10.50, 11.00}\,Hz. (c)The enhanced magnetic sensitivity reaches 22.3\,fT/Hz$^{1/2}$ at resonance frequency $10.00$\,Hz. }
	\label{sen}
\end{figure}

\emph{Spin sensor}.
To search for these pseudomagnetic fields, we use a spin-based amplifier~\cite{jiang2021search,su2021search,jiang2021Floquet,SI}, 
which employs spatially overlapping ensembles of spin-polarized $^{87}$Rb and $^{129}$Xe~\cite{jiang2019floquet}, as shown in Fig.\,\ref{sen}(a).
Hyperpolarized $^{129}$Xe nuclear spins act as an amplifier for the resonant signal from pseudomagnetic fields and the $^{87}$Rb spins function as a conventional magnetometer to detect the enhanced field.
This technique takes advantage of the nuclear magnetic resonance between the modulated pseudomagnetic field $\textbf{B}_{a}$ and $^{129}$Xe spins.
When the oscillation frequency of pseudomagnetic fields matches the $^{129}$Xe Larmor frequency, the $^{129}$Xe spins are slightly tilted away from their polarization axis and generate an oscillating magnetization~\cite{SI,jiang2021search,su2021search}.
These spins act as a transducer converting the pseudomagnetic field into the effective magnetic field probed with $^{87}$Rb spins.
Benefiting from the Fermi-contact interactions between $^{129}$Xe spins and $^{87}$Rb spins, the induced $^{129}$Xe transverse magnetization produces an amplified effective magnetic field $|\textbf{B}_{\textrm{eff}}| \gg |\textbf{B}_{a}|$ on $^{87}$Rb spins~\cite{jiang2021search,su2021search}, as shown in Fig.\,\ref{sen}. 
Thus, the spin-based amplifier allows achieving high sensitivity, making it suitable to resonantly search for pseudomagnetic fields generated by $V_{pp}$.
Here, we ignore the pseudomagnetic field along $\hat{z}$ because the spin-based amplifier is insensitive to oscillating fields along this direction.

The amplification effect can be quantitively described by the amplification factor~\cite{jiang2021search,su2021search,SI,jiang2021Floquet}
\begin{equation}
     \label{eff}
    \eta=|\textbf{B}_{\textrm{eff}}|/|\textbf{B}_{\textrm{a}}| = \dfrac{4 \pi}{3} \kappa_0 M^{n}_0 P^{n}_{0} \gamma_{n} T_{2n},
\end{equation}
where $\kappa_0$ is the Fermi-contact factor; $M^{n}_0$ is the maximum magnetization of $^{129}$Xe nuclei associated with full spin polarization; ${P}^{n}_0$ is the equilibrium polarization of $^{129}$Xe nuclei; $\gamma_n$ is the gyromagnetic radio of the $^{129}$Xe nucleus; and $T_{2n}$ is the transverse relaxation time of $^{129}$Xe spins.
As seen in Eq.\,\eqref{eff}, considerable amplification requires long transverse relaxation times, high vapor density, and high polarization of nuclear spins.
The experiments reported here are performed with a spin-based amplifier similar to that of Refs.\,\cite{jiang2021search,su2021search,jiang2021Floquet}, depicted in Fig.\,\ref{setup}(a).
The $^{129}$Xe Larmor frequency is tuned to 10.00\,Hz by setting the bias field along $\hat{z}$ at 847\,nT.
To calibrate the system, an additional oscillating field of 13.0\,pT is applied along $\hat{y}$. 
Scanning the oscillation frequency of this field, we find that the spin-based amplifier responds as a single-pole band-pass filter~\cite{jiang2021search,su2021search} with a full-width at half-peak amplitude of 49\,mHz.
Calibration of the amplification factor is performed at several different frequencies between 9.00\,Hz and 11.00\,Hz with the averaged amplification factor $\eta \approx 43.5\pm 0.8$, as shown in Fig.\,\ref{sen}(b). 
Therefore, the magnetic sensitivity is enhanced to $\approx$22.3\,fT/$\textrm{Hz}^{1/2}$, whereas the off-resonance sensitivity of $^{87}$Rb magnetometer is only about 1.0\,pT/$\textrm{Hz}^{1/2}$, as shown in Fig.\,\ref{sen}(c). 

We note that both spin-based amplifiers and self-compensating comagnetometers \cite{lee2018improved,vasilakis2009limits,almasi2020new} use overlapping spin ensembles (e.g., $^{129}$Xe-$^{87}$Rb),
whereas they are quite different in their frequency response~\cite{SI}. 
Comagnetometers operate in a specific bias field, where $^{129}$Xe spins and $^{87}$Rb spins are strongly coupled~\cite{padniuk2021self}.
The normal magnetic field is cancelled by the $^{129}$Xe magnetization and the $^{87}$Rb spins are then in zero-field environment, leaving comagnetometers sensitive to low-frequency exotic fields (e.g., the one modulated at 0.18\,Hz in Ref.\,\cite{vasilakis2009limits}).
In contrast, the spin-based amplifier operates in a broad range of bias fields, where the $^{129}$Xe spins and $^{87}$Rb
are weakly coupled.
The effective field generated by the $^{129}$Xe magnetization is measured with $^{87}$Rb spins, resulting in the sensitivity to pseudomagnetic fields with oscillation frequency above 1\,Hz.
Therefore, the spin-based amplifier is well suited to searching for new physics predicted by numerous theories beyond the Standard Model, for example, ultralight axion-like dark matter~\cite{jiang2021search}.

\begin{figure}[t]  
	\makeatletter
	\def\@captype{figure}
	\makeatother
	\includegraphics[scale=1.05]{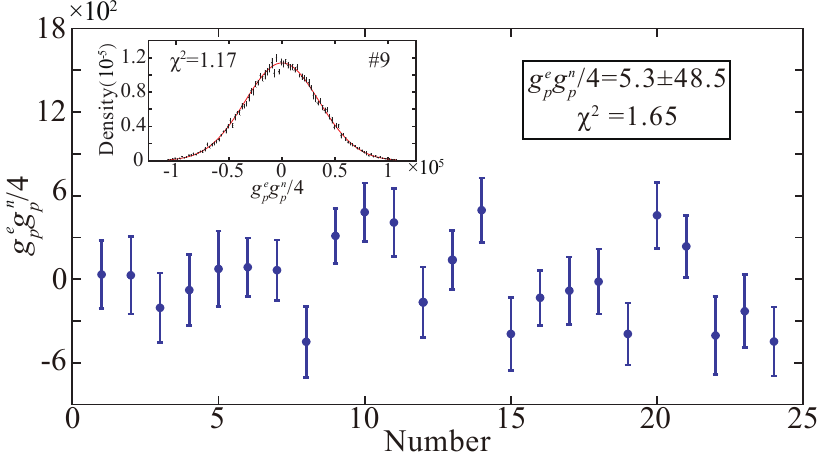}
	\caption{Experimental coupling strength at $m_a=0.1$~meV. Each point and its error bar represent the average of one hour and its corresponding standard error over approximately one hour, respectively.
	The weighted reduced chi-square is $\chi^2=1.65$. Inset top: Graph of values for each 3600 sec-long record closely follows a Gaussian distribution for 9th hour data with $\chi^2 = 1.17$.}
	\label{M1}
\end{figure}

\emph{Search experiments and data analysis}.~
Due to the proximity of the spin source to the spin sensor,
the source produces a measurable magnetic dipole field on the sensor.
The dipole field is calibrated with 
a commercial miniaturized atomic magnetometer (from QuSpin Inc.) at the position of the spin-sensor cell 1.
The magnetic dipole field is measured to be about 3.2\,pT with the QuSpin magnetometer, in good agreement with the finite element analysis of magnetic field~\cite{SI}.
To eliminate the magnetic dipole field,
the spin source is shielded with two-layer $\mu$-metal shields and the spin sensor is shielded with a one-layer $\mu$-metal shield, as shown in Fig.\,\ref{setup}(a).
The total shielding factor is experimentally determined to be $>$$10^6$, and thus the dipole field is reduced to $<3.2$\,aT,
which is negligible with respect to the magnetic sensitivity in our experiment.


During the experiment, the spin-based amplifier is tuned to resonance frequency to match the optical chopping frequency of the spin-source pump laser, i.e., $\nu_0 \approx \nu \approx 10.00$\,Hz.
Due to the narrow bandwidth (49\,mHz) of the amplifier, only the first harmonic of the pseudomagnetic fields is amplified [see Eq.\,\eqref{B3}] and the effect of other harmonics is negligible~\cite{SI}.
The data were collected in six-hour batches, after which the spin-based amplifier was readjusted to optimize the magnetic-field sensitivity by tweaking the bias field, etc.
While recording search data,
the parameters of the spin source were monitored, such as the chopper frequency and pump power.
The total duration of the search experiment was about 24 hours.

In data analysis, a ``lock-in'' scheme is used to extract the weak signal with a known carrier frequency from noisy environment~\cite{ji2018new,su2021search}.
The reference frequency for this procedure is given by the chopper frequency, and the phase is measured by applying an auxiliary oscillating magnetic field with a coil. 
The weak signal of the spin-based amplifier is separated into segments of a single period $\approx 0.1$~s and a corresponding experimental coupling strength $|g_p^e g_p^n|/4$ is extracted.
A histogram of such coupling strength and a corresponding Gaussian fit is obtained to provide the mean value and standard error for each record (1 hour), as shown in Fig.~\ref{M1}.
The final coupling strength is obtained as $|g_p^e g_p^n|/4 \approx (5.3\pm48.5_{\text{stat}})$ for $m_a=0.1$~meV with a weighted reduced chi-square $\chi^2 = 1.65$.
The details are presented in Supplementary Material (section V)~\cite{SI}.

\begin{table}[t]
\newcommand{\tabincell}[2]{\begin{tabular}{@{}#1@{}}#2\end{tabular}}
\begin{ruledtabular}
\caption {~~~Summary of corrections to $|g_p^e g_p^n|/4$ (presented here for $m_a =0.1$\,meV) and systematic errors.} 
\label{tab:my_label}
\renewcommand{\arraystretch}{1.2}
\begin{tabular}{l c c}   
Parameter & Value & $\Delta g_p^e g_p^n/4$ 
  \\[0.1cm]
\hline 
Position of cell 2 $x$ (mm) & $1.9\pm0.2$ & ${}^{<0.01}_{-0.04}$  \\[0.15cm]
Position of cell 2 $y$ (mm) & $-23.1\pm1.0$ & ${}^{-0.98}_{+1.23}$  \\ [0.15cm]
Position of cell 2 $z$ (mm) & $41.9\pm0.3$ & ${}^{+0.75}_{-0.69}$  \\ [0.15cm]
Num. of polarized Rb ($10^{14}$) & $2.18\pm0.18$ & ${}^{-0.41}_{+0.44}$ \\[0.15cm]
Phase delay $\phi$ (deg) & $102.6\pm0.6$ & ${}^{-1.45}_{+1.45}$\\[0.15cm]
Calib. const. $\alpha$ (V/nT) & $2.11^{+0.002}_{-0.386}$ & ${}^{-0.01}_{+1.20}$\\[0.25cm]

Final $|g_p^e g_p^n|/4$ & $5.3$
 & $\pm 48.5 \ (\text{statistical}) $\\

$(m_a =0.1~\text{meV})$ &  & $\pm 2.4 \ (\text{systematic})
$ 
\end{tabular} 
\end{ruledtabular}
\end{table}

Systematic errors are summarized in Table~\ref{tab:my_label}. 
It was found that one of important contributions to the uncertainty comes from the phase variation of the reference signal after the re-optimization. 
Another important contribution is from the calibration constant $\alpha$, the coefficient transferring the output signal (V) of the spin-based amplifier to the unknown pseudomagnetic field (nT). 
The fluctuation of $\alpha$ can be caused by the intrinsic instability of the spin-based amplifier including the fluctuation of the laser beam and temperature and the instability of the operation frequency $\nu$ of the chopper.
The number of polarized $^{87}$Rb spins and its corresponding standard error $(2.18\pm0.18)\times 10^{14}$ is characterized by the parameters of the spin source, such as the optical absorption, the atomic density and the size of the cell 2.
The overall systematic uncertainty is derived by combining all the systematic errors in quadrature~\cite{SI}. 
Accordingly, we quote the final total coupling strength $|g_p^e g_p^n|/4$ as $(5.3\pm48.5_{\text{stat}}\pm2.4_{\text{syst}})$ at $m_a=0.1$~meV.

\begin{figure}[t]  
	\makeatletter
	\def\@captype{figure}
	\makeatother
	\includegraphics[scale=1.015]{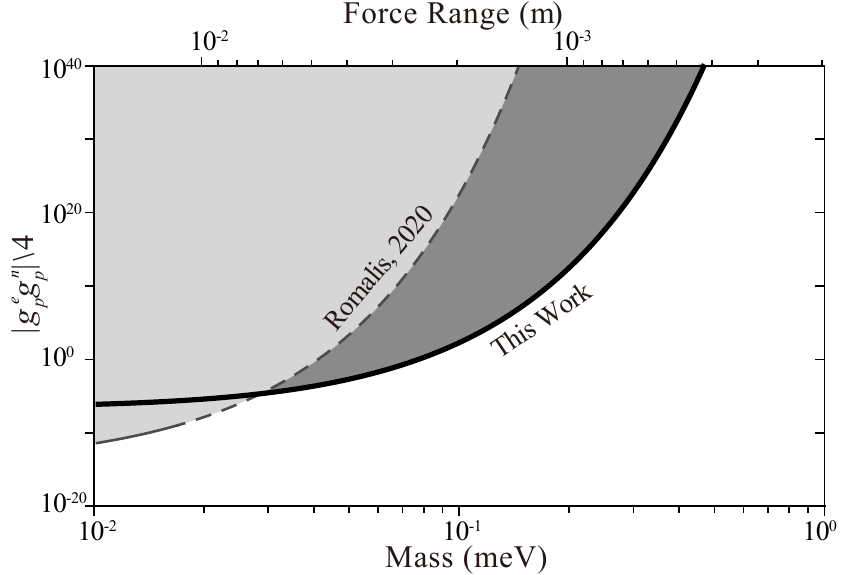}
	\caption{Experimental constraints on $|g_p^e g_p^n|/4$ within the axion window. The thick dark curve represents the experimental limit on the electron-neutron coupling with 1.96$\sigma$ uncertainty as a function of the boson mass $m_a$ in the bottom axis and the force range $\lambda \!= \!\hbar(m_a c)^{-1}$ in the top axis. The thin gray curve indicates the same coupling derived from comagnetometer experiments~\cite{almasi2020new}, while the dashed one is deduced based on Eq.~\ref{v3}. Our limit sets new constraint on axions or axionlike particles for $m_a>0.03$~meV within the axion window.} 
\end{figure}

Figure 4 shows the new constraints on $|g_p^e g_p^n|/4$ set by this work together with recent constraint from experimental searches for exotic dipole-dipole interactions within the axion window~\cite{almasi2020new}. 
The solid line represents the constraint of our work, corresponding to 1.96 times the quadrature sum of the statistical error and systematic error (95$\%$ confidence level).
The deduced constraints on $|g_p^e g_p^n|/4$ combination were derived from comagnetometer experiments presented as the dashed line~\cite{almasi2020new}.
For the mass range from 0.03~meV to 1~meV, our result provides the most stringent constraint from direct measurements on $|g_p^e g_p^n|/4$.

In conclusion, we have reported new direct constraints,
based on a spin-based amplifier, on exotic spin-spin interactions within the well-motivated axion window.
Although demonstrated for the axion searches, the spin-based amplifier can be generically applied into resonant searches for other new particles beyond the Standard Model, for example, spin-1 bosons such as paraphotons and Z' bosons \cite{dobrescu2005massless,su2021search} or dark photon~\cite{an2015direct,chaudhuri2015radio}.
Spin-1 bosons can mediate exotic spin-spin velocity-dependent interactions ($V_{6+7}$, $V_{8}$, $V_{14}$, $V_{15}$ and $V_{16}$).
In this case, the resonantly oscillating pseudomagnetic fields can be generated by modulating the relative speed between the source and $^{129}$Xe amplifier.
Combining our current $^{129}$Xe based amplifier and recently developed SmCo$_5$ spin sources~\cite{ji2018new, almasi2020new}, 
the sensitivity to $|f_{6+7}| < 10^{-20}, |f_{14}| < 10^{-31}, |f_{15}| < 10^{-4}, |f_{16}| < 10^{-4}, |f_{8}| < 10^{-23}$ can potentially be achieved and would provide the most stringent constraints on the new light boson for the force range from 0.01\,m to 10\,m.
Moreover, the spin-based amplifier can be used to explore exotic interactions beyond the Standard Model
present in the ambient space, such as the ``axion-field gradient''~\cite{jiang2021search}.
A further improvement of the experimental sensitivity to exotic interactions can be anticipated by employing $^3$He-based amplifier, which has much longer spin-coherence time than that of $^{129}$Xe~\cite{vasilakis2009limits}, allowing one to improve sensitivity to those exotic interactions by several orders of magnitude~\cite{jiang2021search,su2021search}.

~\

This work was supported by National Key Research and Development Program of China (grant no. 2018YFA0306600), National Natural Science Foundation of China (grants nos. 11661161018, 11927811, 12004371), Anhui Initiative in Quantum Information Technologies (grant no. AHY050000),
USTC Research Funds of the Double First-Class Initiative (grant no. YD3540002002). This work was also supported by the Cluster of Excellence ``Precision Physics, Fundamental Interactions, and Structure of Matter'' (PRISMA+ EXC 2118/1) funded by the German Research Foundation (DFG) within the German Excellence Strategy (Project ID 39083149).

\bibliographystyle{naturemag}
\bibliography{mainrefs}

\begin{thebibliography}{10}
\expandafter\ifx\csname url\endcsname\relax
  \def\url#1{\texttt{#1}}\fi
\expandafter\ifx\csname urlprefix\endcsname\relax\def\urlprefix{URL }\fi
\providecommand{\bibinfo}[2]{#2}
\providecommand{\eprint}[2][]{\url{#2}}

\bibitem{demille2017probing}
\bibinfo{author}{DeMille, D.}, \bibinfo{author}{Doyle, J.~M.} \&
  \bibinfo{author}{Sushkov, A.~O.}
\newblock \bibinfo{title}{Probing the frontiers of particle physics with
  tabletop-scale experiments}.
\newblock \emph{\bibinfo{journal}{Science}} \textbf{\bibinfo{volume}{357}},
  \bibinfo{pages}{990--994} (\bibinfo{year}{2017}).

\bibitem{ramsey1979tensor}
\bibinfo{author}{Ramsey, N.~F.}
\newblock \bibinfo{title}{The tensor force between two protons at long range}.
\newblock \emph{\bibinfo{journal}{Physica (Amsterdam)}}
  \textbf{\bibinfo{volume}{96}}, \bibinfo{pages}{285--289}
  (\bibinfo{year}{1979}).

\bibitem{wilczek1978problem}
\bibinfo{author}{Wilczek, F.}
\newblock \bibinfo{title}{Problem of strong \rm{P} and \rm{T} invariance in the
  presence of instantons}.
\newblock \emph{\bibinfo{journal}{Phys. Rev. Lett.}}
  \textbf{\bibinfo{volume}{40}}, \bibinfo{pages}{279} (\bibinfo{year}{1978}).

\bibitem{peccei1977cp}
\bibinfo{author}{Peccei, R.~D.} \& \bibinfo{author}{Quinn, H.~R.}
\newblock \bibinfo{title}{\rm{CP} conservation in the presence of
  pseudoparticles}.
\newblock \emph{\bibinfo{journal}{Phys. Rev. Lett.}}
  \textbf{\bibinfo{volume}{38}}, \bibinfo{pages}{1440} (\bibinfo{year}{1977}).

\bibitem{weinberg1978new}
\bibinfo{author}{Weinberg, S.}
\newblock \bibinfo{title}{A new light boson?}
\newblock \emph{\bibinfo{journal}{Phys. Rev. Lett.}}
  \textbf{\bibinfo{volume}{40}}, \bibinfo{pages}{223} (\bibinfo{year}{1978}).

\bibitem{fadeev2019revisiting}
\bibinfo{author}{Fadeev, P.} \emph{et~al.}
\newblock \bibinfo{title}{Revisiting spin-dependent forces mediated by new
  bosons: Potentials in the coordinate-space representation for macroscopic-and
  atomic-scale experiments}.
\newblock \emph{\bibinfo{journal}{Phys. Rev. A}} \textbf{\bibinfo{volume}{99}},
  \bibinfo{pages}{022113} (\bibinfo{year}{2019}).

\bibitem{safronova2018search}
\bibinfo{author}{Safronova, M.} \emph{et~al.}
\newblock \bibinfo{title}{Search for new physics with atoms and molecules}.
\newblock \emph{\bibinfo{journal}{Rev. Mod. Phys.}}
  \textbf{\bibinfo{volume}{90}}, \bibinfo{pages}{025008}
  (\bibinfo{year}{2018}).

\bibitem{kim2019experimental}
\bibinfo{author}{Kim, Y.~J.}, \bibinfo{author}{Chu, P.-H.},
  \bibinfo{author}{Savukov, I.} \& \bibinfo{author}{Newman, S.}
\newblock \bibinfo{title}{Experimental limit on an exotic parity-odd spin-and
  velocity-dependent interaction using an optically polarized vapor}.
\newblock \emph{\bibinfo{journal}{Nat. Commun.}} \textbf{\bibinfo{volume}{10}},
  \bibinfo{pages}{1--7} (\bibinfo{year}{2019}).

\bibitem{yan2013new}
\bibinfo{author}{Yan, H.} \& \bibinfo{author}{Snow, W.}
\newblock \bibinfo{title}{New limit on possible long-range parity-odd
  interactions of the neutron from neutron-spin rotation in liquid
  $^4$\rm{He}}.
\newblock \emph{\bibinfo{journal}{Phys. Rev. Lett.}}
  \textbf{\bibinfo{volume}{110}}, \bibinfo{pages}{082003}
  (\bibinfo{year}{2013}).

\bibitem{jiang2021search}
\bibinfo{author}{Jiang, M.}, \bibinfo{author}{Su, H.}, \bibinfo{author}{Garcon,
  A.}, \bibinfo{author}{Peng, X.} \& \bibinfo{author}{Budker, D.}
\newblock \bibinfo{title}{Search for axion-like dark matter with spin-based
  amplifiers}.
\newblock \emph{\bibinfo{journal}{Nat. Phys.}} \textbf{\bibinfo{volume}{17}},
  \bibinfo{pages}{1402--1407} (\bibinfo{year}{2021}).

\bibitem{jaeckel2010low}
\bibinfo{author}{Jaeckel, J.} \& \bibinfo{author}{Ringwald, A.}
\newblock \bibinfo{title}{The low-energy frontier of particle physics}.
\newblock \emph{\bibinfo{journal}{Annu. Rev. Nucl. Part Sci}}
  \textbf{\bibinfo{volume}{60}}, \bibinfo{pages}{405--437}
  (\bibinfo{year}{2010}).

\bibitem{langacker2009physics}
\bibinfo{author}{Langacker, P.}
\newblock \bibinfo{title}{The physics of heavy z′ gauge bosons}.
\newblock \emph{\bibinfo{journal}{Rev. Mod. Phys.}}
  \textbf{\bibinfo{volume}{81}}, \bibinfo{pages}{1199} (\bibinfo{year}{2009}).

\bibitem{kim2010axions}
\bibinfo{author}{Kim, J.~E.} \& \bibinfo{author}{Carosi, G.}
\newblock \bibinfo{title}{Axions and the strong \rm{CP} problem}.
\newblock \emph{\bibinfo{journal}{Rev. Mod. Phys.}}
  \textbf{\bibinfo{volume}{82}}, \bibinfo{pages}{557} (\bibinfo{year}{2010}).

\bibitem{irastorza2018new}
\bibinfo{author}{Irastorza, I.~G.} \& \bibinfo{author}{Redondo, J.}
\newblock \bibinfo{title}{New experimental approaches in the search for
  axion-like particles}.
\newblock \emph{\bibinfo{journal}{Prog. Part. Nucl. Phys.}}
  \textbf{\bibinfo{volume}{102}}, \bibinfo{pages}{89--159}
  (\bibinfo{year}{2018}).

\bibitem{svrcek2006axions}
\bibinfo{author}{Svrcek, P.} \& \bibinfo{author}{Witten, E.}
\newblock \bibinfo{title}{Axions in string theory}.
\newblock \emph{\bibinfo{journal}{J. High Energy Phys.}}
  \textbf{\bibinfo{volume}{2006}}, \bibinfo{pages}{051} (\bibinfo{year}{2006}).

\bibitem{kamionkowski2014dark}
\bibinfo{author}{Kamionkowski, M.}, \bibinfo{author}{Pradler, J.} \&
  \bibinfo{author}{Walker, D.~G.}
\newblock \bibinfo{title}{Dark energy from the string axiverse}.
\newblock \emph{\bibinfo{journal}{Phys. Rev. Lett.}}
  \textbf{\bibinfo{volume}{113}}, \bibinfo{pages}{251302}
  (\bibinfo{year}{2014}).

\bibitem{afach2021search}
\bibinfo{author}{Afach, S.} \emph{et~al.}
\newblock \bibinfo{title}{Search for topological defect dark matter using the
  global network of optical magnetometers for exotic physics searches
  (\rm{GNOME})}.
\newblock \emph{\bibinfo{journal}{Nat. Phys.}} \textbf{\bibinfo{volume}{17}},
  \bibinfo{pages}{1396--1401} (\bibinfo{year}{2021}).

\bibitem{turner1990windows}
\bibinfo{author}{Turner, M.~S.}
\newblock \bibinfo{title}{Windows on the axion}.
\newblock \emph{\bibinfo{journal}{Phys. Rep.}} \textbf{\bibinfo{volume}{197}},
  \bibinfo{pages}{67--97} (\bibinfo{year}{1990}).

\bibitem{youdin1996limits}
\bibinfo{author}{Youdin, A.}, \bibinfo{author}{Krause~Jr, D.},
  \bibinfo{author}{Jagannathan, K.}, \bibinfo{author}{Hunter, L.} \&
  \bibinfo{author}{Lamoreaux, S.}
\newblock \bibinfo{title}{Limits on spin-mass couplings within the axion
  window}.
\newblock \emph{\bibinfo{journal}{Phys. Rev. Lett.}}
  \textbf{\bibinfo{volume}{77}}, \bibinfo{pages}{2170} (\bibinfo{year}{1996}).

\bibitem{arvanitaki2014resonantly}
\bibinfo{author}{Arvanitaki, A.} \& \bibinfo{author}{Geraci, A.~A.}
\newblock \bibinfo{title}{Resonantly detecting axion-mediated forces with
  nuclear magnetic resonance}.
\newblock \emph{\bibinfo{journal}{Phys. Rev. Lett.}}
  \textbf{\bibinfo{volume}{113}}, \bibinfo{pages}{161801}
  (\bibinfo{year}{2014}).

\bibitem{almasi2020new}
\bibinfo{author}{Almasi, A.}, \bibinfo{author}{Lee, J.},
  \bibinfo{author}{Winarto, H.}, \bibinfo{author}{Smiciklas, M.} \&
  \bibinfo{author}{Romalis, M.~V.}
\newblock \bibinfo{title}{New limits on anomalous spin-spin interactions}.
\newblock \emph{\bibinfo{journal}{Phys. Rev. Lett.}}
  \textbf{\bibinfo{volume}{125}}, \bibinfo{pages}{201802}
  (\bibinfo{year}{2020}).

\bibitem{ji2018new}
\bibinfo{author}{Ji, W.} \emph{et~al.}
\newblock \bibinfo{title}{New experimental limits on exotic
  spin-spin-velocity-dependent interactions by using \rm{SmCo}$_5$ spin
  sources}.
\newblock \emph{\bibinfo{journal}{Phys. Rev. Lett.}}
  \textbf{\bibinfo{volume}{121}}, \bibinfo{pages}{261803}
  (\bibinfo{year}{2018}).

\bibitem{jiang2020interference}
\bibinfo{author}{Jiang, M.} \emph{et~al.}
\newblock \bibinfo{title}{Interference in atomic magnetometry}.
\newblock \emph{\bibinfo{journal}{Adv. Quantum Technol.}}
  \textbf{\bibinfo{volume}{3}}, \bibinfo{pages}{2000078}
  (\bibinfo{year}{2020}).

\bibitem{bulatowicz2013laboratory}
\bibinfo{author}{Bulatowicz, M.} \emph{et~al.}
\newblock \bibinfo{title}{Laboratory search for a long-range \rm{T}-odd,
  \rm{P}-odd interaction from axionlike particles using dual-species nuclear
  magnetic resonance with polarized $^{129}$\rm{Xe} and $^{131}$\rm{Xe} gas}.
\newblock \emph{\bibinfo{journal}{Phys. Rev. Lett.}}
  \textbf{\bibinfo{volume}{111}}, \bibinfo{pages}{102001}
  (\bibinfo{year}{2013}).

\bibitem{aggarwal2020characterization}
\bibinfo{author}{Aggarwal, N.} \emph{et~al.}
\newblock \bibinfo{title}{Characterization of magnetic field noise in the
  \rm{ARIADNE} source mass rotor}.
\newblock \emph{\bibinfo{journal}{arXiv:2011.12617}}  (\bibinfo{year}{2020}).

\bibitem{su2021search}
\bibinfo{author}{Su, H.} \emph{et~al.}
\newblock \bibinfo{title}{Search for exotic spin-dependent interactions with a
  spin-based amplifier}.
\newblock \emph{\bibinfo{journal}{Sci. Adv.}} \textbf{\bibinfo{volume}{7}},
  \bibinfo{pages}{eabi9535} (\bibinfo{year}{2021}).

\bibitem{heckel2013limits}
\bibinfo{author}{Heckel, B.}, \bibinfo{author}{Terrano, W.} \&
  \bibinfo{author}{Adelberger, E.}
\newblock \bibinfo{title}{Limits on exotic long-range spin-spin interactions of
  electrons}.
\newblock \emph{\bibinfo{journal}{Phys. Rev. Lett.}}
  \textbf{\bibinfo{volume}{111}}, \bibinfo{pages}{151802}
  (\bibinfo{year}{2013}).

\bibitem{terrano2015short}
\bibinfo{author}{Terrano, W.}, \bibinfo{author}{Adelberger, E.},
  \bibinfo{author}{Lee, J.} \& \bibinfo{author}{Heckel, B.}
\newblock \bibinfo{title}{Short-range, spin-dependent interactions of
  electrons: A probe for exotic pseudo-goldstone bosons}.
\newblock \emph{\bibinfo{journal}{Phys. Rev. Lett.}}
  \textbf{\bibinfo{volume}{115}}, \bibinfo{pages}{201801}
  (\bibinfo{year}{2015}).

\bibitem{wineland1991search}
\bibinfo{author}{Wineland, D.}, \bibinfo{author}{Bollinger, J.},
  \bibinfo{author}{Heinzen, D.}, \bibinfo{author}{Itano, W.} \&
  \bibinfo{author}{Raizen, M.}
\newblock \bibinfo{title}{Search for anomalous spin-dependent forces using
  stored-ion spectroscopy}.
\newblock \emph{\bibinfo{journal}{Phys. Rev. Lett.}}
  \textbf{\bibinfo{volume}{67}}, \bibinfo{pages}{1735} (\bibinfo{year}{1991}).

\bibitem{rong2018searching}
\bibinfo{author}{Rong, X.} \emph{et~al.}
\newblock \bibinfo{title}{Searching for an exotic spin-dependent interaction
  with a single electron-spin quantum sensor}.
\newblock \emph{\bibinfo{journal}{Nat. Commun.}} \textbf{\bibinfo{volume}{9}},
  \bibinfo{pages}{1--7} (\bibinfo{year}{2018}).

\bibitem{hunter2013using}
\bibinfo{author}{Hunter, L.}, \bibinfo{author}{Gordon, J.},
  \bibinfo{author}{Peck, S.}, \bibinfo{author}{Ang, D.} \&
  \bibinfo{author}{Lin, J.-F.}
\newblock \bibinfo{title}{Using the earth as a polarized electron source to
  search for long-range spin-spin interactions}.
\newblock \emph{\bibinfo{journal}{Science}} \textbf{\bibinfo{volume}{339}},
  \bibinfo{pages}{928--932} (\bibinfo{year}{2013}).

\bibitem{hunter2014using}
\bibinfo{author}{Hunter, L.} \& \bibinfo{author}{Ang, D.}
\newblock \bibinfo{title}{Using geoelectrons to search for velocity-dependent
  spin-spin interactions}.
\newblock \emph{\bibinfo{journal}{Phys. Rev. Lett.}}
  \textbf{\bibinfo{volume}{112}}, \bibinfo{pages}{091803}
  (\bibinfo{year}{2014}).

\bibitem{glenday2008limits}
\bibinfo{author}{Glenday, A.~G.}, \bibinfo{author}{Cramer, C.~E.},
  \bibinfo{author}{Phillips, D.~F.} \& \bibinfo{author}{Walsworth, R.~L.}
\newblock \bibinfo{title}{Limits on anomalous spin-spin couplings between
  neutrons}.
\newblock \emph{\bibinfo{journal}{Phys. Rev. Lett.}}
  \textbf{\bibinfo{volume}{101}}, \bibinfo{pages}{261801}
  (\bibinfo{year}{2008}).

\bibitem{jiang2019floquet}
\bibinfo{author}{Jiang, M.}, \bibinfo{author}{Su, H.}, \bibinfo{author}{Wu,
  Z.}, \bibinfo{author}{Peng, X.} \& \bibinfo{author}{Budker, D.}
\newblock \bibinfo{title}{Floquet maser}.
\newblock \emph{\bibinfo{journal}{Sci. Adv.}} \textbf{\bibinfo{volume}{7}},
  \bibinfo{pages}{eabe0719} (\bibinfo{year}{2021}).

\bibitem{ding2020constraints}
\bibinfo{author}{Ding, J.} \emph{et~al.}
\newblock \bibinfo{title}{Constraints on the velocity and spin dependent exotic
  interaction at the micrometer range}.
\newblock \emph{\bibinfo{journal}{Phys. Rev. Lett.}}
  \textbf{\bibinfo{volume}{124}}, \bibinfo{pages}{161801}
  (\bibinfo{year}{2020}).

\bibitem{dobrescu2006spin}
\bibinfo{author}{Dobrescu, B.~A.} \& \bibinfo{author}{Mocioiu, I.}
\newblock \bibinfo{title}{Spin-dependent macroscopic forces from new particle
  exchange}.
\newblock \emph{\bibinfo{journal}{J. High Energy Phys.}}
  \textbf{\bibinfo{volume}{2006}}, \bibinfo{pages}{005} (\bibinfo{year}{2006}).

\bibitem{jiang2019magnetic}
\bibinfo{author}{Jiang, M.} \emph{et~al.}
\newblock \bibinfo{title}{Magnetic gradiometer for the detection of zero-to
  ultralow-field nuclear magnetic resonance}.
\newblock \emph{\bibinfo{journal}{Phys. Rev. Appl.}}
  \textbf{\bibinfo{volume}{11}}, \bibinfo{pages}{024005}
  (\bibinfo{year}{2019}).

\bibitem{an2015direct}
\bibinfo{author}{An, H.}, \bibinfo{author}{Pospelov, M.},
  \bibinfo{author}{Pradler, J.} \& \bibinfo{author}{Ritz, A.}
\newblock \bibinfo{title}{Direct detection constraints on dark photon dark
  matter}.
\newblock \emph{\bibinfo{journal}{Phys. Lett. B}}
  \textbf{\bibinfo{volume}{747}}, \bibinfo{pages}{331--338}
  (\bibinfo{year}{2015}).

\bibitem{SI}
\bibinfo{title}{Supplementary material}  (\bibinfo{year}{2021}).

\bibitem{walker1997spin}
\bibinfo{author}{Walker, T.~G.} \& \bibinfo{author}{Happer, W.}
\newblock \bibinfo{title}{Spin-exchange optical pumping of noble-gas nuclei}.
\newblock \emph{\bibinfo{journal}{Rev. Mod. Phys.}}
  \textbf{\bibinfo{volume}{69}}, \bibinfo{pages}{629} (\bibinfo{year}{1997}).

\bibitem{ni1999search}
\bibinfo{author}{Ni, W.-T.}, \bibinfo{author}{Pan, S.-s.},
  \bibinfo{author}{Yeh, H.-C.}, \bibinfo{author}{Hou, L.-S.} \&
  \bibinfo{author}{Wan, J.}
\newblock \bibinfo{title}{Search for an axionlike spin coupling using a
  paramagnetic salt with a dc \rm{SQUID}}.
\newblock \emph{\bibinfo{journal}{Phys. Rev. Lett.}}
  \textbf{\bibinfo{volume}{82}}, \bibinfo{pages}{2439} (\bibinfo{year}{1999}).

\bibitem{kimball2015nuclear}
\bibinfo{author}{Kimball, D.~J.}
\newblock \bibinfo{title}{Nuclear spin content and constraints on exotic
  spin-dependent couplings}.
\newblock \emph{\bibinfo{journal}{New J. Phys.}} \textbf{\bibinfo{volume}{17}},
  \bibinfo{pages}{073008} (\bibinfo{year}{2015}).

\bibitem{jiang2021Floquet}
\bibinfo{author}{Jiang, M.} \emph{et~al.}
\newblock \bibinfo{title}{Floquet spin amplification}.
\newblock \emph{\bibinfo{journal}{arXiv:2112.06190}}  (\bibinfo{year}{2021}).

\bibitem{lee2018improved}
\bibinfo{author}{Lee, J.}, \bibinfo{author}{Almasi, A.} \&
  \bibinfo{author}{Romalis, M.}
\newblock \bibinfo{title}{Improved limits on spin-mass interactions}.
\newblock \emph{\bibinfo{journal}{Phys. Rev. Lett.}}
  \textbf{\bibinfo{volume}{120}}, \bibinfo{pages}{161801}
  (\bibinfo{year}{2018}).

\bibitem{vasilakis2009limits}
\bibinfo{author}{Vasilakis, G.}, \bibinfo{author}{Brown, J.},
  \bibinfo{author}{Kornack, T.} \& \bibinfo{author}{Romalis, M.}
\newblock \bibinfo{title}{Limits on new long range nuclear spin-dependent
  forces set with a \textrm{K}-$^3$\textrm{He} comagnetometer}.
\newblock \emph{\bibinfo{journal}{Phys. Rev. Lett.}}
  \textbf{\bibinfo{volume}{103}}, \bibinfo{pages}{261801}
  (\bibinfo{year}{2009}).

\bibitem{padniuk2021self}
\bibinfo{author}{Padniuk, M.} \emph{et~al.}
\newblock \bibinfo{title}{Self-compensating co-magnetometer vs. spin-exchange
  relaxation-free magnetometer: sensitivity to nonmagnetic spin couplings}.
\newblock \emph{\bibinfo{journal}{arXiv:2107.05501}}  (\bibinfo{year}{2021}).

\bibitem{dobrescu2005massless}
\bibinfo{author}{Dobrescu, B.~A.}
\newblock \bibinfo{title}{Massless gauge bosons other than the photon}.
\newblock \emph{\bibinfo{journal}{Phys. Rev. Lett.}}
  \textbf{\bibinfo{volume}{94}}, \bibinfo{pages}{151802}
  (\bibinfo{year}{2005}).

\bibitem{chaudhuri2015radio}
\bibinfo{author}{Chaudhuri, S.} \emph{et~al.}
\newblock \bibinfo{title}{Radio for hidden-photon dark matter detection}.
\newblock \emph{\bibinfo{journal}{Phys. Rev. D}} \textbf{\bibinfo{volume}{92}},
  \bibinfo{pages}{075012} (\bibinfo{year}{2015}).

\end{thebibliography}


\begin{thebibliography}{1}
\expandafter\ifx\csname url\endcsname\relax
  \def\url#1{\texttt{#1}}\fi
\expandafter\ifx\csname urlprefix\endcsname\relax\def\urlprefix{URL }\fi
\providecommand{\bibinfo}[2]{#2}
\providecommand{\eprint}[2][]{\url{#2}}

\bibitem{jiang2021search}
\bibinfo{author}{Jiang, M.}, \bibinfo{author}{Su, H.}, \bibinfo{author}{Garcon,
  A.}, \bibinfo{author}{Peng, X.} \& \bibinfo{author}{Budker, D.}
\newblock \bibinfo{title}{Search for axion-like dark matter with spin-based
  amplifiers}.
\newblock \emph{\bibinfo{journal}{Nat. Phys.}} \textbf{\bibinfo{volume}{17}},
  \bibinfo{pages}{1402--1407} (\bibinfo{year}{2021}).

\bibitem{su2021search}
\bibinfo{author}{Su, H.} \emph{et~al.}
\newblock \bibinfo{title}{Search for exotic spin-dependent interactions with a
  spin-based amplifier}.
\newblock \emph{\bibinfo{journal}{Sci. Adv.}} \textbf{\bibinfo{volume}{7}},
  \bibinfo{pages}{eabi9535} (\bibinfo{year}{2021}).

\bibitem{fadeev2019revisiting}
\bibinfo{author}{Fadeev, P.} \emph{et~al.}
\newblock \bibinfo{title}{Revisiting spin-dependent forces mediated by new
  bosons: Potentials in the coordinate-space representation for macroscopic-and
  atomic-scale experiments}.
\newblock \emph{\bibinfo{journal}{Phys. Rev. A}} \textbf{\bibinfo{volume}{99}},
  \bibinfo{pages}{022113} (\bibinfo{year}{2019}).

\bibitem{dobrescu2006spin}
\bibinfo{author}{Dobrescu, B.~A.} \& \bibinfo{author}{Mocioiu, I.}
\newblock \bibinfo{title}{Spin-dependent macroscopic forces from new particle
  exchange}.
\newblock \emph{\bibinfo{journal}{J. High Energy Phys.}}
  \textbf{\bibinfo{volume}{2006}}, \bibinfo{pages}{005} (\bibinfo{year}{2006}).

\bibitem{walker1997spin}
\bibinfo{author}{Walker, T.~G.} \& \bibinfo{author}{Happer, W.}
\newblock \bibinfo{title}{Spin-exchange optical pumping of noble-gas nuclei}.
\newblock \emph{\bibinfo{journal}{Rev. Mod. Phys.}}
  \textbf{\bibinfo{volume}{69}}, \bibinfo{pages}{629} (\bibinfo{year}{1997}).

\bibitem{Romalis1997Pressure}
\bibinfo{author}{Romalis, M.~V.}, \bibinfo{author}{Miron, E.} \&
  \bibinfo{author}{Cates, G.~D.}
\newblock \bibinfo{title}{Pressure broadening of rb ${D}_{1}$ and ${D}_{2}$
  lines by ${}^{3} \text{He}$, ${}^{4} \text{He}$, $\text{N}_{2}$, and
  $\text{Xe}$: Line cores and near wings}.
\newblock \emph{\bibinfo{journal}{Phys. Rev. A}} \textbf{\bibinfo{volume}{56}},
  \bibinfo{pages}{4569--4578} (\bibinfo{year}{1997}).

\bibitem{ji2018new}
\bibinfo{author}{Ji, W.} \emph{et~al.}
\newblock \bibinfo{title}{New experimental limits on exotic
  spin-spin-velocity-dependent interactions by using \rm{SmCo}$_5$ spin
  sources}.
\newblock \emph{\bibinfo{journal}{Phys. Rev. Lett.}}
  \textbf{\bibinfo{volume}{121}}, \bibinfo{pages}{261803}
  (\bibinfo{year}{2018}).

\bibitem{lee2018improved}
\bibinfo{author}{Lee, J.}, \bibinfo{author}{Almasi, A.} \&
  \bibinfo{author}{Romalis, M.}
\newblock \bibinfo{title}{Improved limits on spin-mass interactions}.
\newblock \emph{\bibinfo{journal}{Phys. Rev. Lett.}}
  \textbf{\bibinfo{volume}{120}}, \bibinfo{pages}{161801}
  (\bibinfo{year}{2018}).

\end{thebibliography}

\end{document}


\title{Supplementary materials for: \\``Limits on axions and axionlike particles within the axion window using a spin-based amplifier''}

\date{\today}

\author{Yuanhong Wang}
\email[]{These authors contributed equally to this work}
\affiliation{
CAS Key Laboratory of Microscale Magnetic Resonance and School of Physical Sciences, University of Science and Technology of China, Hefei, Anhui 230026, China}
\affiliation{
CAS Center for Excellence in Quantum Information and Quantum Physics, University of Science and Technology of China, Hefei, Anhui 230026, China}

\author{Haowen Su}
\email[]{These authors contributed equally to this work}
\affiliation{
CAS Key Laboratory of Microscale Magnetic Resonance and School of Physical Sciences, University of Science and Technology of China, Hefei, Anhui 230026, China}
\affiliation{
CAS Center for Excellence in Quantum Information and Quantum Physics, University of Science and Technology of China, Hefei, Anhui 230026, China}

\author{Min Jiang}
\email[]{dxjm@ustc.edu.cn}
\affiliation{
CAS Key Laboratory of Microscale Magnetic Resonance and School of Physical Sciences, University of Science and Technology of China, Hefei, Anhui 230026, China}
\affiliation{
CAS Center for Excellence in Quantum Information and Quantum Physics, University of Science and Technology of China, Hefei, Anhui 230026, China}

\author{Ying Huan}
\affiliation{
CAS Key Laboratory of Microscale Magnetic Resonance and School of Physical Sciences, University of Science and Technology of China, Hefei, Anhui 230026, China}
\affiliation{
CAS Center for Excellence in Quantum Information and Quantum Physics, University of Science and Technology of China, Hefei, Anhui 230026, China}

\author{\mbox{Yushu Qin}}
\affiliation{
CAS Key Laboratory of Microscale Magnetic Resonance and School of Physical Sciences, University of Science and Technology of China, Hefei, Anhui 230026, China}
\affiliation{
CAS Center for Excellence in Quantum Information and Quantum Physics, University of Science and Technology of China, Hefei, Anhui 230026, China}

\author{\mbox{Chang Guo}}
\affiliation{
CAS Key Laboratory of Microscale Magnetic Resonance and School of Physical Sciences, University of Science and Technology of China, Hefei, Anhui 230026, China}
\affiliation{
CAS Center for Excellence in Quantum Information and Quantum Physics, University of Science and Technology of China, Hefei, Anhui 230026, China}

\author{\mbox{Zehao Wang}}
\affiliation{
CAS Key Laboratory of Microscale Magnetic Resonance and School of Physical Sciences, University of Science and Technology of China, Hefei, Anhui 230026, China}
\affiliation{
CAS Center for Excellence in Quantum Information and Quantum Physics, University of Science and Technology of China, Hefei, Anhui 230026, China}

\author{Dongdong Hu}
\affiliation{
State Key Laboratory of Particle Detection and Electronics, University of Science and Technology of China, Hefei, Anhui 230026, China}

\author{Wei Ji}
\affiliation{Helmholtz-Institut, GSI Helmholtzzentrum f{\"u}r Schwerionenforschung, Mainz 55128, Germany}
\affiliation{Johannes Gutenberg University, Mainz 55128, Germany}

\author{Pavel~Fadeev}
\affiliation{Helmholtz-Institut, GSI Helmholtzzentrum f{\"u}r Schwerionenforschung, Mainz 55128, Germany}
\affiliation{Johannes Gutenberg University, Mainz 55128, Germany}

\author{Xinhua Peng}
\email[]{xhpeng@ustc.edu.cn}
\affiliation{
CAS Key Laboratory of Microscale Magnetic Resonance and School of Physical Sciences, University of Science and Technology of China, Hefei, Anhui 230026, China}
\affiliation{
CAS Center for Excellence in Quantum Information and Quantum Physics, University of Science and Technology of China, Hefei, Anhui 230026, China}

\author{Dmitry Budker}
\affiliation{Helmholtz-Institut, GSI Helmholtzzentrum f{\"u}r Schwerionenforschung, Mainz 55128, Germany}
\affiliation{Johannes Gutenberg University, Mainz 55128, Germany}
\affiliation{Department of Physics, University of California, Berkeley, CA 94720-7300, USA}

\maketitle

\tableofcontents

\section{Spin sensor}
\label{Ssen}

In this section,
we describe the details of spin-based amplifier (spin sensor),
which has been recently demonstrated in Refs.~\cite{jiang2021search,su2021search}.
In this work,
we use a spin-based amplifier to search for the exotic spin-dependent interaction $V_{pp}$ described in Ref.~\cite{fadeev2019revisiting} (equivalently $V_3$ in Ref.~\cite{dobrescu2006spin}).
As shown in Sec.~\ref{sec31},
the exotic interaction $V_{pp}$ can generate an oscillating pseudomagnetic field on the spin-based amplifier and can be amplified by it by a factor of more than 40.

\subsection{Schematic of the spin-based amplifier}
\label{secA1}

\begin{figure} [h] 
	\makeatletter
	\def\@captype{figure}
	\makeatother
	\includegraphics[scale=1.20]{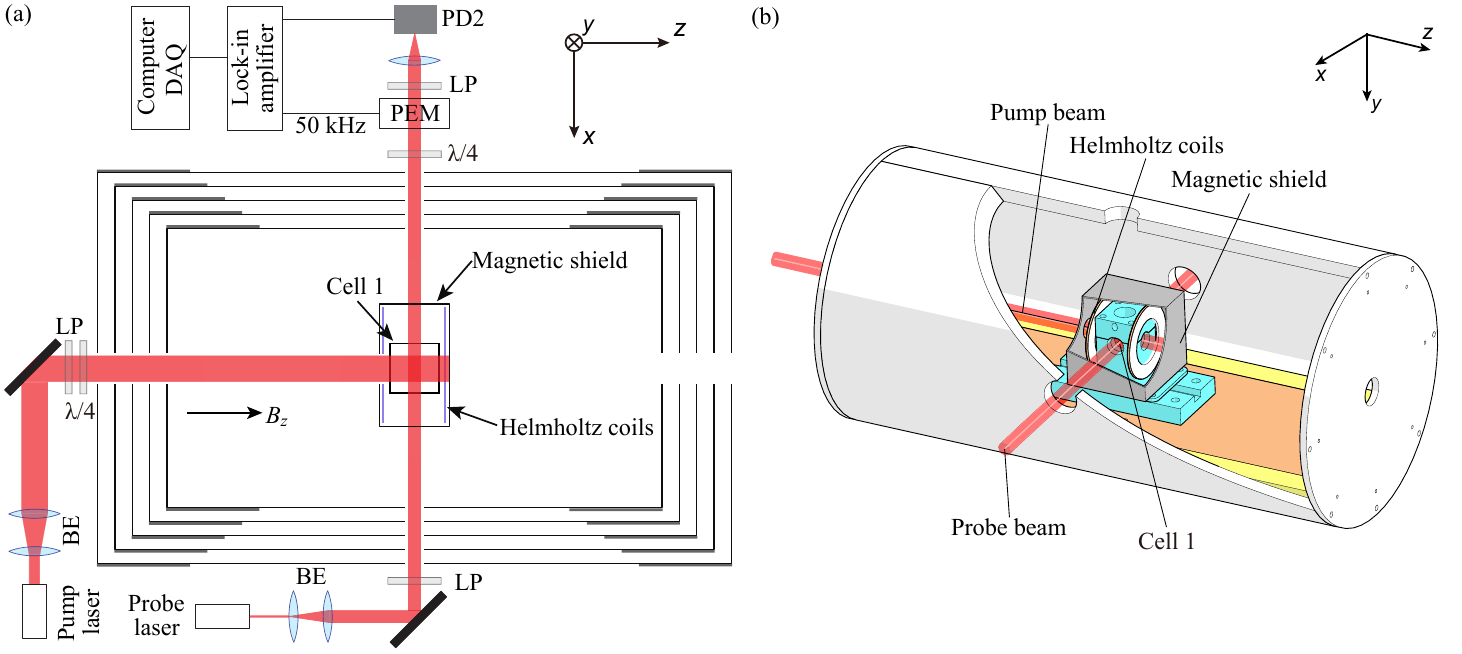}
	\caption{\textbf{Experimental setup of the spin sensor}. (a) Two-dimensional diagram of the spin-based amplifier as spin sensor. A vapor cell (cell 1) contains spatially overlapping ensembles of $^{129}$Xe and $^{87}$Rb and is heated to 165$^\circ$C. A circularly polarized pump laser polarizes $^{87}$Rb atoms along $z$. $^{129}$Xe spins are polarized with polarized $^{87}$Rb atoms through spin-exchange collisions. A resonant pseudomagnetic field can be amplified by $^{129}$Xe spins, which produce an effective magnetic field $\textbf{B}_{\textrm{eff}}^n$ measured $\emph{in situ}$ with $^{87}$Rb atoms. BE, beam expander; LP, linear polarizer; $\lambda /4$, quarter-wave plate; PD, photodiode; PEM, photoelastic modulator; DAQ, data acquisition. (b) Three-dimensional diagram of the spin-based amplifier.}
	\label{setup}
\end{figure}

The details of the experimental apparatus are shown in Fig.~\ref{setup}.
The spin-based amplifier uses a 0.5~cm$^3$ cubic vapor cell (cell 1) containing 5~torr of isotopically enriched $^{129}$Xe,
250~torr N$_2$ as buffer gas,
and a droplet (several milligrams) of isotopically enriched $^{87}$Rb.
$^{87}$Rb atoms are polarized by a circularly polarized laser at 795~nm and probed by a linearly polarized laser blue-detuned 110~GHz from the D2 transition at 780 nm.
$^{129}$Xe spins are polarized by spin-exchange collisions with polarized $^{87}$Rb atoms.
In order to match the oscillation frequency $\nu$ of the pseudomagnetic field generated by the exotic interaction $V_{pp}$ (see the details of spin source in Sec.~\ref{Ss} and the pseudomagnetic field in Sec.~\ref{sec3}),
the $^{129}$Xe Larmor frequency $\nu_0$ is tuned to $\nu$ by adjusting the bias magnetic field $B_z^0$ applied with a set of Helmholtz coils [see Fig.~\ref{setup}(b)].
In the resonant case ($\nu_0 \approx \nu$),
the polarized $^{129}$Xe spins are tilted away from the $z$ axis,
and then generate oscillating transverse magnetization.
Due to the Fermi-contact interactions between $^{87}$Rb and $^{129}$Xe spins,
the $^{129}$Xe transverse magnetization produces an effective oscillating field on $^{87}$Rb atoms,
which functions as a magnetometer to \emph{in situ} measure this field~\cite{jiang2021search,su2021search}.
In particular, we experimentally demonstrate that the effective field strength is much larger than the pseudomagnetic field strength by a factor of about 40 (see the main text).
As a result,
$^{129}$Xe spins act as a transducer converting the pseudomagnetic field into the effective magnetic field probed with $^{87}$Rb spins.
We present the experimental calibration for the spin-based amplifier in Sec.~\ref{secC1}.

\subsection{Analysis of spin-based amplifier}

The exotic spin-spin interactions $V_{pp}$ mediated by axions can produce an pseudomagnetic field (see Sec.~\ref{sec3}).
In the following, we present the theoretical analysis of the response of the spin-based amplifier to this field.
We also perform experimental calibrations by applying an oscillating field to simulate the pseudomagnetic field.

The Fermi-contact interaction between them introduces an effective magnetic field~\cite{walker1997spin}
\begin{equation}
\textbf{B}^{e,n}_{\textrm{eff}}=\beta M_0^{e,n}\textbf{P}^{e,n},
\end{equation}
where $\textbf{B}^{e}_{\textrm{eff}}$ ($\textbf{B}^{n}_{\textrm{eff}}$) represents the effective magnetic field experienced by $^{129}$Xe ($^{87}$Rb) spins,
$\beta = 8\pi \kappa_0 /3$,
$\kappa_0$ is the enhancement factor of the Fermi-contact interaction,
$M_0^{e}$ ($M_0^{n}$) is the maximum magnetization of $^{87}$Rb electron ($^{129}$Xe nucleus), and $\textbf{P}^{e}$ ($\textbf{P}^{n}$) is the polarization of $^{87}$Rb electron ($^{129}$Xe nucleus).
In contrast to the "self-compensating" comagnetometers where $^{87}$Rb and $^{129}$Xe spins are strongly coupled with each other,
the spin-based amplifier operates under a relatively larger bias field $B_z^{0}$,
leading to the weak coupling between $^{87}$Rb and $^{129}$Xe spins.
The effective field $\textbf{B}^{e}_{\textrm{eff}}$ of electron spins is similar as a weak static field along $z$ that slightly shifts the $^{129}$Xe Larmor frequency.
The effective field $\textbf{B}^{n}_{\textrm{eff}}$ of nuclear spins with the enhancement of the pseudomagnetic field is measured $\emph{in situ}$ with the $^{87}$Rb spins.

The resonant response of the spin-based amplifier to an oscillating magnetic field, for example, $\textbf{B}_{\textrm{ac}}^{\textrm{exo}}=B_{\textrm{ac}}^{\textrm{exo}} \cos(2 \pi \nu t)\hat{\bm{y}}$, can be described with the Bloch equation
\begin{equation}
    \frac{\partial \textbf{P}^{n}}{\partial t}=\gamma_{n}({B}_z\hat{\bm{z}}+\textbf{B}_{\textrm{ac}}^{\textrm{exo}}) \times \textbf{P}^{n} + \frac{P^{n}_{0}\hat{\bm{z}} -\textbf{P}^{n} }{\{T_{2n},T_{2n},T_{1n}\}},
    \label{A1}
\end{equation}
where $\gamma_n$ is the gyromagnetic ratio of $^{129}$Xe nucleus,
${B}_z$ is total field of the applied bias field and $^{87}$Rb effective magnetic field,
$P^n_0$ is the equilibrium polarization of $^{129}$Xe nucleus,
and $T_{1n}$ ($T_{2n}$) is the longitudinal (transverse) relaxation time of $^{129}$Xe spins. Based on Eq.~(\ref{A1}),
the steady-state solution of $^{129}$Xe spins polarization $\textbf{P}^n$ is~\cite{jiang2021search,su2021search}
\begin{align}
\label{E6}
\textbf{P}^n &=\dfrac{1}{2}P^{n}_{0} \gamma_{n} B_{\textrm{ac}}^{\textrm{exo}}\dfrac{ T_{2n}\cos(2\pi\nu t)+2\pi(\nu-\nu_{0})T_{2n}^{2}\sin(2\pi\nu t)}{1+({\gamma_{n} B_{\textrm{ac}}^{\textrm{exo}}}/2)^{2}T_{1n}T_{2n}+[2\pi(\nu-\nu_{0})]^{2}T_{2n}^{2}}\hat{\bm{x}},\\
\label{E7}
&+\dfrac{1}{2}P^{n}_{0} \gamma_{n} B_{\textrm{ac}}^{\textrm{exo}}\dfrac{T_{2n}\sin(2\pi\nu t)-2\pi(\nu-\nu_{0})T_{2n}^{2}\cos(2\pi\nu t)}{1+({\gamma_{n} B_{\textrm{ac}}^{\textrm{exo}}}/2)^{2}T_{1n}T_{2n}+[2\pi(\nu-\nu_{0})]^{2}T_{2n}^{2}}\hat{\bm{y}},\\
\label{E8}
&+P^n_0 \frac{1+[2\pi(\nu-\nu_{0})]^{2}T_{2n}^{2}}{1+({\gamma_{n} B_{\textrm{ac}}^{\textrm{exo}}}/2)^{2}T_{1n}T_{2n}+[2\pi(\nu-\nu_{0})]^{2}T_{2n}^{2}}\hat{\bm{z}}.
\end{align}
It should be noted that the pseudomagnetic field is weak enough to satisfy $({\gamma_{n} B_{\textrm{ac}}^{\textrm{exo}}}/2)^{2}T_{1n}T_{2n} \ll 1$.
As a result, the longitudinal polarization can be approximated as $P^n_z\approx P_0^n$, which generates an effective static field $\beta M^{n}_0 P^{n}_{0}$ on $^{87}$Rb spins. 
In the case of transverse polarization, according to $\textbf{B}_{\rm{eff}}^n=\beta M^{n}_0 \textbf{P}^{n}$, an oscillating magnetic field is experienced by the $^{87}$Rb spins
\begin{equation}
\begin{aligned}
\label{E9}
&\textbf{B}_{\rm{\rm{eff}}}^n =\overbrace{\dfrac{1}{2}\beta M^{n}_0 P^{n}_{0} \gamma_{n} B_{\textrm{ac}}^{\textrm{exo}}\dfrac{T_{2n}\cos(2\pi\nu t)+2\pi(\nu-\nu_{0})T_{2n}^{2}\sin(2\pi\nu t)}{1+[2\pi(\nu-\nu_{0})]^{2}T_{2n}^{2}}\hat{\bm{x}}}^\textrm{ effective field generated by $^{129}$Xe $x$ magnetization}+\overbrace{\dfrac{1}{2}\beta M^{n}_0 P^{n}_{0} \gamma_{n} B_{\textrm{ac}}^{\textrm{exo}}\dfrac{T_{2n}\sin(2\pi\nu t)-2\pi(\nu-\nu_{0})T_{2n}^{2}\cos(2\pi\nu t)}{1+[2\pi(\nu-\nu_{0})]^{2}T_{2n}^{2}}\hat{\bm{y}}}^\textrm{effective field generated by $^{129}$Xe $y$ magnetization}.
\end{aligned}
\end{equation}
Under the resonance condition $\nu=\nu_0$,
the strength of the effective transverse field achieves a maximum 
\begin{equation}
	 \textbf{B}_{\textrm{eff}}^n(\nu = \nu_0)=\dfrac{4\pi}{3}\kappa_0 M^{n}_0 P^{n}_{0} \gamma_{n} T_{2n}[\cos(2\pi\nu t)\hat{\bm{x}} + \sin(2\pi\nu t)\hat{\bm{y}}]B_{\textrm{ac}}^{\textrm{exo}}.
	 \label{beff}
\end{equation}
Such an effective magnetic field oscillates with an amplitude of $\dfrac{4\pi}{3}\kappa_0 M^{n}_0 P^{n}_{0} \gamma_{n} T_{2n}B_{\textrm{ac}}^{\textrm{exo}}$ in the $xy$ plane and can be measured $\emph{in situ}$ with a $^{87}$Rb magnetometer. 
The response of atomic magnetometers to the effective field and its derivation can be found in Ref.~\cite{jiang2021search,su2021search}.

\subsection*{1. Pseudomagnetic-field amplification}
\label{sec31}

Benefiting from the Fermi-contact interaction,
the signal from $\textbf{B}_{\rm{eff}}^n$ can be greatly larger than that from the oscillating pseudomagnetic field $\textbf{B}^{\rm{exo}}_{\rm{ac}}$.
To quantitatively describe the considerable enhancement, we introduce an amplification factor
\begin{equation}
\eta=|\textbf{B}_{\rm{eff}}^n|/|\textbf{B}^{\textrm{exo}}_{\rm{ac}}|=\dfrac{4\pi}{3}\kappa_0 M^{n}_0 P^{n}_{0} \gamma_{n} T_{2n}.
\label{E12}
\end{equation}
To realize a considerable amplification effect, several experimental parameters $\{\kappa_0,M^{n}_0 ,P^{n}_{0},T_{2n}\}$ need to be optimized.
Specifically, long transverse relaxation times, high vapor density, and high polarization of nuclear spins are required.

Before the search experiments,
the amplification factor is experimentally calibrated by applying an auxiliary oscillating field along $y$.
In the resonant case, for example, the bias field $B_z^0$ is set as $\approx 849.17$~nT to tune the $^{129}$Xe Larmor frequency.
Because of the residual magnetic field and effective field of $^{87}$Rb atoms,
the exact Larmor frequency $\nu_0$ is experimentally calibrated using the $^{129}$Xe free-decay signal.
A resonant oscillating field $\nu \approx 10.00$~Hz is applied along $y$ and the corresponding amplitude of the output signal of $^{87}$Rb magnetometer is recorded as $A_{\textrm{Reson}}$.
In off-resonant case, an oscillating field is applied along $y$ at a frequency of 70.00~Hz (the response is independent of frequency in a broad range) and the amplitude of the $^{87}$Rb magnetometer output signal is recorded as a reference $A_{\textrm{FORes}}$.
By comparing these two amplitudes, we obtain the amplification factor
\begin{equation}
    \eta \approx  \dfrac{A_{\textrm{Reson}}}{A_{\textrm{FORes}}}.
\end{equation}
As shown in Fig.~2(b) in the main text,
the amplification factor is calibrated to be $43.5\pm0.8$ at frequencies {9.00, 9.50, 10.00, 10.50, 11.00}\,Hz.
The enhanced magnetic sensitivity reaches 22.3\,fT/Hz$^{1/2}$ at resonance frequency $10.00$\,Hz, as shown in Fig.~2(c) in the main text.

\subsection*{2. Frequency bandwidth}
\label{sec4}

The amplification factor describes the amplification effect in the resonant case.
Next, we analyse the frequency dependence of the amplification.
Based on Eq.~(\ref{E9}), 
the effective magnetic field $|\textbf{B}_{\textrm{eff}}^n|$ can be rewritten as a function of the frequency difference $\nu-\nu_0$ in the amplitude spectrum,
\begin{equation}
\begin{aligned}
|\textbf{B}_{\textrm{eff}}^n(\nu)| \propto \dfrac{\Lambda/2}{\sqrt{(\nu-\nu_{0})^2+(\Lambda/2)^2}}.
\label{E15}
\end{aligned}
\end{equation}
The full-width at half-maximum (FWHM) is $\sqrt{3}\Lambda$.
Thus, the spin-based amplifier can enhance the near-resonant signal.
The FWHM bandwidth is experimentally calibrated by scanning the frequency of an auxiliary oscillating field along $y$.
For example, we set the bias field as $B_{z}^{0} \approx 849.17$~nT.
An oscillating field with strength $\approx 13.0$~pT is applied along $y$. 
We scan the oscillating-field frequency near the resonance frequency.
The experimental data are fitted according to Eq.~(\ref{E15}),
the bandwidth $\sqrt{3}\Lambda \approx 49$~mHz of the spin-based amplifier is obtained, as shown in Fig.~2(c) in the main text.

\section{Spin source}
\label{Ss}

In this section,
we describe the details of the spin-source setup and calibration experiments,
such as the measurement of spatial distribution of spin polarization and the number of polarized Rb spins in the spin-source vapor cell.

\subsection{Schematic of the spin source}

\begin{figure} [h] 
	\makeatletter
	\def\@captype{figure}
	\makeatother
	\includegraphics[scale=1.4]{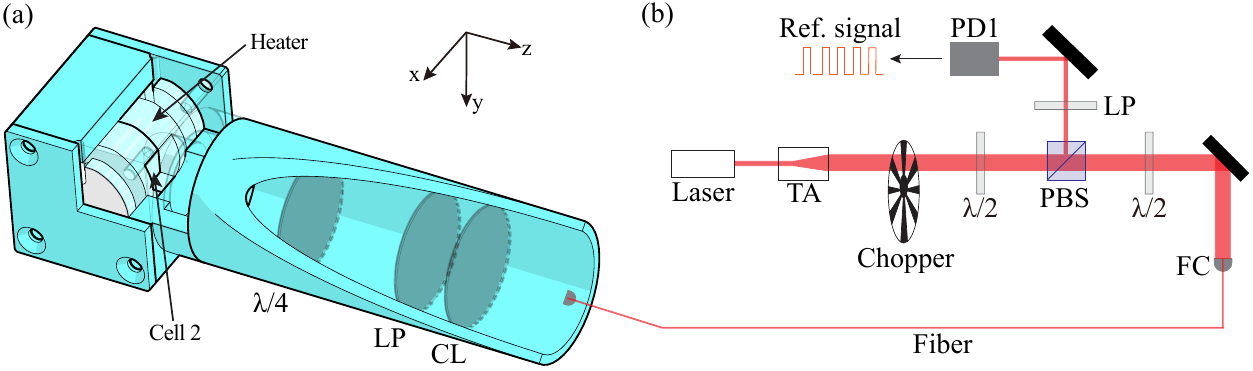}
	\caption{\textbf{Schematic of the spin source}. (a) Spin-source setup containing a $^{87}$Rb vapor cell, a heater, and optical elements. A $5.8\,\text{cm}^3$ cubic vapor cell containing a droplet of $^{87}$Rb and about 0.37\,amg N$_2$ is heated to 180$^\circ$C. The high-optical-power pump beam (0.5\,W) is expanded and circularly polarized. $^{87}$Rb spins are polarized to $\sim$92$\%$ and the number of polarized spins is calculated to be about $2.18\times10^{14}$. (b) High-power pump system. An optical chopper modulates the laser beam with a square-wave pattern, with a frequency of about $10~\rm{Hz}$ and a duty cycle of about $50\%$, thus modulating the spin polarization of $^{87}$Rb atoms. A half-wave plate and a polarizing beamsplitter are used to split the laser beam into two beams with different powers. The stronger one is coupled to a single-mode fiber with an optical fiber coupler. The weaker one is measured with PD1 to monitor the laser power. TA, tapered amplifier; $\lambda$/4, quarter-wave plate; $\lambda$/2, half-wave plate; CL, convex lens; PBS, polarizing beamsplitter; PD, photodiode; LP, linear polarizer; FC, fiber coupler.}
	\label{SI_2_2_0}
\end{figure}

The spin-source setup (see Fig.\,\ref{SI_2_2_0}) consists of a $^{87}$Rb vapor cell (cell 2),
a heater, and a high-power optical pumping system.
The spin-source cell contains a droplet of isotopically enriched $^{87}$Rb and 0.37\,amg $\rm{N_2}$ as buffer gas.
In order to increase the atomic number density of $^{87}$Rb gas,
electric heater coils are used to heat the vapor cell to $180\,^\circ\rm C$.
To reduce the magnetic-field interference originating from the heating current, bifilar winding is used, so the magnetic field is mostly compensated 
and the frequency of the current is chosen to be $110\,\rm{kHz}$, above the frequencies of interest in this experiment.
The heater coils are wrapped on a nonmagnetic thermally conductive boron-nitride cylinder, with the
vapor cell placed in its center, as shown in Fig.\,\ref{SI_2_2_0}(a).
The whole spin-source setup is placed inside an insulated non-magnetic PEEK chamber and the entire heating system is wrapped with aerogel for thermal insulation.

As shown in Fig.~\ref{SI_2_2_0}(b), a high-power pump system containing a high-power pump laser of 795\,nm (generated by a tapered amplifier laser system from TOPTICA Photonics Inc.) is used to polarize $^{87}$Rb atoms in cell 2.
In order to modulate the spin polarization of $^{87}$Rb atoms,
an optical chopper is placed in the laser path to periodically block the pump beam at a frequency $\nu \approx 10.00$\,Hz with 50$\%$ duty cycle.
Then a half-wave plate and a polarizing beamsplitter (PBS) are used to split the pump light into two beams with the power ratio of more than $100:1$.
The weaker beam is monitored in real-time by a photodiode as reference signal and
the stronger one is used for optical pumping.
The second half-wave plate is used to adjust the light polarization of the laser beam for efficiently coupling the laser with a single-mode fiber.
The coupling efficiency of the fiber is approximately $70\%$.
The laser transmitted from the fiber is expanded by a beam expander and adjusted to be circularly polarized by a linear polarizer and a quarter-wave plate.
These optical elements are placed in a cylinder box and are connected to the heating chamber, as shown in Fig.~\ref{SI_2_2_0}(a).

\subsection{Calibration of spin-source parameters}
\label{SsB}

The magnitude of the pseudomagnetic field generated by the spin source is proportional to the number of the polarized electron spins in the spin-source cell.
Due to the optical absorption and spatial distribution of pump-beam intensity,
the spin polarization is not uniform over the spin-source cell 2.
Therefore, the pseudomagnetic field is dependent on the spatial spin distribution $\rho(\bm{r})$ in the spin source [see Eq.~(\ref{B3})],
\begin{equation}
\rho(\bm{r})=n_{\rm{Rb}}\cdot P_z(\bm{r}),
\label{rho}
\end{equation}
where $n_\text{Rb}$ is the number density of $^{87}$Rb and $P_z(\bm{r})$ is the spin polarization of $^{87}$Rb electrons at the position of $\bm{r}=(x,y,z)$.
To obtain the spatial distribution of polarized electron spins,
it is necessary to obtain the spatial distribution $\rho(\bm{r})$.
In the following, we explain how to calibrate the key parameters $n_{\rm{Rb}}$ and $P_z(\bm{r})$.

\subsection*{1. Atomic absorption spectroscopy measurement}

We use
atomic absorption spectroscopy to calibrate the parameters of the optical pumping, such as the number density of $^{87}$Rb atoms, collisionally shifted resonance frequency of D1 transition and the bandwidth caused by pressure broadening. 
A linearly polarized laser beam (the absorption-spectroscopy laser with small light power) with frequency tuned near the Rb D1 transition, propagating along $z$, travels through the spin-source cell 2.
The absorption of such a linearly polarized laser is described by
\begin{equation}
\label{Is0}
    I_{\rm{out}}=I_{\rm{in}}e^{-n_{\rm{Rb}}\sigma(\nu_l)L},
\end{equation}
where $I_{\rm{out}}$ is the transmitted light intensity, $I_{\rm{in}}$ is the incident light intensity, $\sigma(\nu_l)$ is the photon absorption cross-section, $\nu_l$ is the optical frequency of the absorption-spectroscopy laser,
and $L$ is the length of spin-source cell 2. 
Due to the large pressure broadening by the buffer gas N$_2$,
the absorption cross-section can be described as a function of the laser frequency $\nu_l$ with a Lorentzian lineshape~\cite{Romalis1997Pressure}
\begin{equation}
    \sigma(\nu_l)=\pi r_ecf_{\rm{D1}}\frac{\Gamma/2\pi}{(\nu_l-\nu_{\text{D1}})^2+(\Gamma/2)^2},
\label{ab1}
\end{equation}
where $r_e$ is the classical radius of the electron, $c$ is the velocity of light, $f_{\rm{D1}}=1/3$ is the oscillator strength of the D1 transition, $\Gamma$ is the full-width at half-maximum (FWHM) of the lineshape mainly caused by N$_2$,
and $\nu_{\text{D1}}$ is the collisionally shifted resonance frequency of D1 transition. 

In the experiment,
due to the unavoidable reflection from the vapor-cell glass and the window of the oven,
the measured transmitted light intensity is reduced and described by $I'_{\rm{out}}=\xi I_{\rm{out}}$.
Based on Eqs.\,\eqref{ab1} and \eqref{Is0},
optical absorption of the linearly polarized laser light 
can be written as
\begin{equation}
\ln{\frac{I'_{\rm{out}}}{\xi I_{\rm{in}}}}=-\pi r_e c f_{\rm{D1}}n_{\rm{Rb}}L\frac{\Gamma/2\pi}{(\nu_l-\nu_{\text{D1}})^2+(\Gamma/2)^2}.
\label{ab}
\end{equation}

\begin{figure} [h] 
	\makeatletter
	\def\@captype{figure}
	\makeatother
	\includegraphics[scale=1]{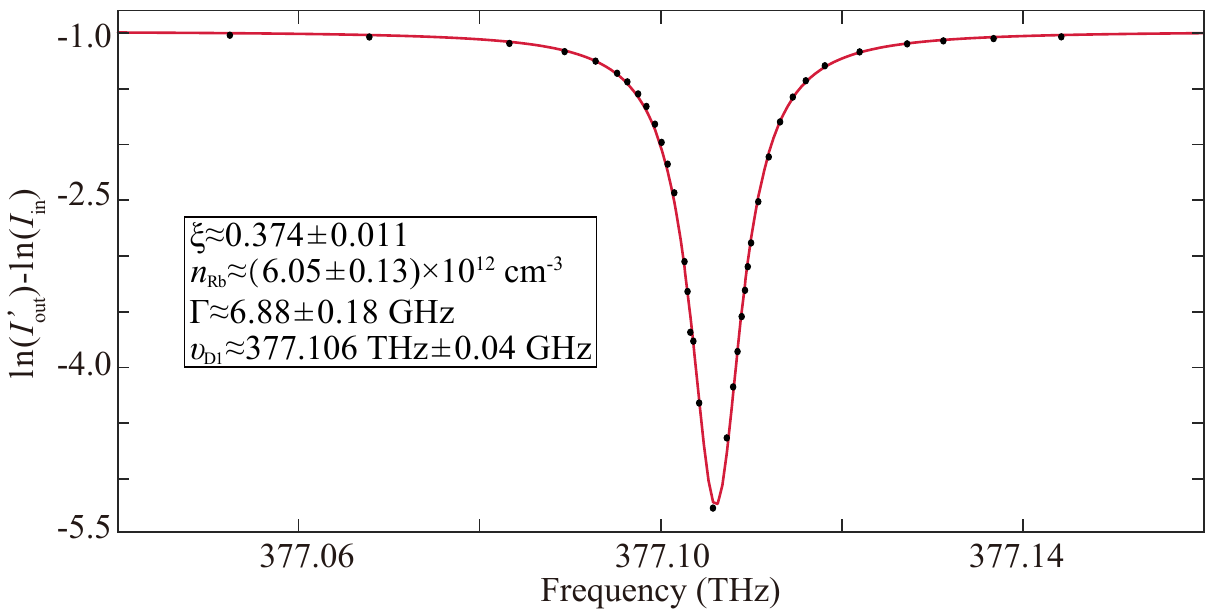}
	\caption{\textbf{Rb atomic absorption spectrum at 100$^\circ$C as an example}. The parameters of the Lorentzian lineshape are fitted to $\xi \approx 0.374\pm0.011$,  $n_{\rm{Rb}} \approx (6.05\pm0.13)\times10^{12}~\rm{cm^{-3}}$, $\Gamma \approx 6.88\pm0.18 \,\rm{GHz}$ and $\nu_{\text{D1}} \approx 377.106\,\rm{THz}\pm0.04\,\text{GHz}$.
    As the temperature rises, number density of $^{87}$Rb increases and the corresponding optical absorption increases, rendering
    the transmitted light intensity near the resonance frequency hard to measure.
    Therefore, parameters independent of temperature, such as $\Gamma$ and $\nu_{\text{D1}}$, are determined at low temperature (100$^\circ$C). 
    }
	\label{SI_2_2_2}
\end{figure}

By detuning the laser frequency $\nu_l$,
we measure the corresponding laser power of the incident light $I_{\rm{in}}$ and that of the transmitted light $I'_{\rm{out}}$,
and then obtain a profile.
By fitting the profile,
we obtain the key parameters $\xi$, $n_\text{Rb}$, $\Gamma$ and $\nu_{\text{D1}}$.
We show an example of 100$^\circ$C in Fig.\,\ref{SI_2_2_2},
where $\xi$, $n_\text{Rb}$, $\Gamma$, and $\nu_{\text{D1}}$ are obtained.
Using the same method for our final search experiment,
we calibrate the $^{87}$Rb number density as
\begin{equation}
    n_{\text{Rb}}\approx(4.0\pm0.2)\times10^{14}~ \text{cm}^{-3}.
\end{equation}

\subsection*{2.~Calibration of the spin polarization}



The spin polarization of $^{87}$Rb electrons and its spatial distribution depend on the parameters of optical pumping.
In the following, we determine the spin polarization based on calibrated optical-pumping parameters.

\begin{itemize}
\item[(1)]\textbf{Overview of the calibration method}

In our experiments,
we use high-power circularly polarized laser light tuned near the Rb D1 transition to pump the spin-source cell 2.
Due to the light absorption along $z$,
the pumping rate and the corresponding spin polarization of Rb atoms are reduced along the light-propagation direction.
The spin polarization can be generally expressed as
\begin{equation}
    \label{Pz1}
    P_z(\bm{r})=\frac{sR_\text{OP}(\bm{r})}{R_\text{OP}(\bm{r})+R_\text{rel}}, 
\end{equation}
where $R_\text{OP}(\bm{r})$ is the pumping rate and $R_\text{rel}$ is the spin relaxation rate,
$s=\pm1$ is the photon spin component along the pumping direction.
$s=-1$ corresponds to $\sigma^-$ light
and $s=+1$ corresponds to $\sigma^+$ light.
Here $R_\text{OP}(\bm{r})$ is position-dependent and $R_\text{rel}$ is position-independent.

The spin-relaxation mechanisms include collisions with buffer-gas atoms, other alkali atoms and the wall of the vapor cell.
In our experiments, by measuring the resonance linewidth at different pump powers, we determined 
the relaxation rate $R_{\rm{rel}}$ to be  
\begin{equation}
    R_{\text{rel}}\approx 524\pm18~\rm{s}^{-1}.
\end{equation}
Because of the use of a strong laser ($\approx 0.5$\,W) to pump cell 2,
the pumping rate $R_\text{OP}(\bm{r})$ is much larger than the relaxation rate $R_{\text{rel}}$.
As a result, the uncertainty of $R_{\text{rel}}$ contributes little to the uncertainty of the total number of polarized atoms as shown in Table \ref{P}.

The pumping rate is the dominant factor influencing the polarization.
Because the linewidth of the pump laser is much narrower than pressure-broadened linewidth of the atomic D1 transition ($\Gamma\approx 6.88\,$GHz),
the incident pump light can be considered monochromatic.
In this case,
$R_\text{OP}$ is approximated as
\begin{equation}
      \label{Rop}
      R_\text{OP}(\bm{r})\approx\sigma(\nu_l)\cdot\Phi(\bm{r}),
\end{equation}
where $\sigma(\nu_l)$ is the photon-absorption cross-section of Rb D1 transition measured with atomic absorption spectroscopy [see Eq.\,\eqref{ab1}],
$\Phi(\bm{r})$ is the total flux of photons of frequency $\nu_l$ (in photons per unit area per unit time).
$\Phi(\bm{r})$ can be expressed as a function of the intensity of the pump laser
\begin{equation}
\label{phi}
\Phi({\bm{r}})=I({\bm{r}})/h\nu_l,
\end{equation}
where $h\nu_l$ is the energy of an incident photon and $I({\bm{r}})$ is the light intensity.
Due to the atomic absorption and cross-sectional shape of the incident light, $I({\bm{r}})$ depends on position.
In the following, we analyse the spatial dependence of the incident light $I({\bm{r}})$.
\end{itemize}

\begin{itemize}
\item[(2)]\textbf{Distribution of the incident light intensity in the $xy$ plane}

The incident light is a Gaussian beam,
whose light intensity is nonuniform in the $xy$ plane.
Because the spin source (cell 2) is placed in the center of the beam,
the distribution of the incident light intensity in the $xy$ plane can be written as (the direction and origin of the coordinate system are shown in the Fig.~\ref{power})
\begin{equation}
I_{\text{in}}(x,y)=I(x,y,z=0)=\frac{2p}{\pi\omega^2}\text{exp}\left[-\frac{2(x^2+y^2)}{\omega^2}\right],
\label{Iin}
\end{equation}
where $p$ is the laser power and $\omega$ is the Gaussian beam waist radius.
To obtain the spatial distribution of the incident light intensity,
the incident light power $p$ and the spot radius of the expanded beam $\omega$ are required.
The spot radius of the expanded beam is measured with a laser beam profiler (from Ophir-Spiricon Inc.),
\begin{equation}
\omega\approx 2.99\pm0.01~\rm{mm}.
\end{equation}
The total power of the pump laser $p$ is measured at the position 1 (see Fig.\,\ref{power}) with an integrating sphere to be
\begin{equation}
p\approx 0.50\pm0.05\,\rm{W}.
\end{equation}
Monitoring the laser power over a month,
we find that the power variation does not exceed $10\%$ without any additional stabilization.

\begin{figure} [h] 
	\makeatletter
	\def\@captype{figure}
	\makeatother
	\includegraphics[scale=1.2]{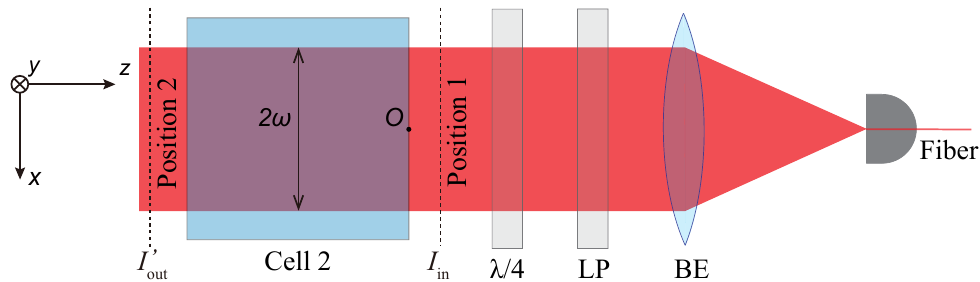}
	\caption{\textbf{Calibration of the incident light intensity in the $xy$ plane.} By measuring the power $p$ and the beam waist radius $\omega$ of the pump laser at the position 1, we calibrate the distribution of the incident light intensity in the $xy$ plane. O, origin; BE, beam expander; LP, linear polarizer; $\lambda /4$, quarter-wave plate.}
	\label{power}
\end{figure}
\end{itemize}

\begin{itemize}
\item[(3)]\textbf{Distribution of the light intensity along $z$}

On- or near-resonant light propagating through the vapor cell can be partially or completely absorbed by the alkali vapor. 
In order to obtain the distribution of the light intensity along $z$,
we consider the attenuation of the circularly polarized light absorbed by the alkali metal atom
\begin{equation}
\frac{\rm d}{\rm{d}z}I(\bm{r})=-n_{\rm{Rb}}\sigma(\nu_l)I(\bm{r})\left [1-sP_z(\bm{r}) \right ].
\label{dI}
\end{equation}
Based on Eqs.\,\eqref{Pz1}, \eqref{Rop} and \eqref{phi}, the differential equation~\eqref{dI} can be reduced to that of one variable $I(\bm{r})$
\begin{equation}
\frac{\rm d}{\rm{d}z}I(\bm{r})=-n_{\rm{Rb}}\sigma(\nu_l)I(\bm{r})\left [1-\frac{\sigma(\nu_l)I(\bm{r})}{\sigma(\nu_l)I(\bm{r})+h\nu_l R_{\rm{rel}}} \right].
\label{IW}
\end{equation}
Equation\,(\ref{IW}) at a fixed $(x,y)$ can be further simplified as~\cite{walker1997spin}
\begin{equation}
    I(z)\cdot\text{exp}\left(\frac{\sigma(\nu_l)I(z)}{h\nu_l R_{\text{rel}}}\right)=I_{\text{in}}\cdot\text{exp}\left [\frac{\sigma(\nu_l)I_{\text{in}}}{h\nu_l R_{\rm{rel}}}-n_{\rm{Rb}}\sigma(\nu_l)|z|\right ].
    \label{Wf}
\end{equation}
To solve the above equation,
we use the Lambert W-function,
which is the inverse function of $f(W)=We^W$.
Equation\,\eqref{Wf} can be rewritten as the form of $We^W$
\begin{equation}
    \overbrace{\frac{\sigma(\nu_l)I(z)}{h\nu_l R_{\text{rel}}}}^{W}\cdot \overbrace{\text{exp}\left(\frac{\sigma(\nu_l)I(z)}{h\nu_l R_{\text{rel}}}\right)}^{e^W}=\overbrace{\frac{\sigma(\nu_l)I_{\text{in}}}{h\nu_l R_{\text{rel}}}\cdot\text{exp}\left [\frac{\sigma(\nu_l)I_{\text{in}}}{h\nu_l R_{\rm{rel}}}-n_{\rm{Rb}}\sigma(\nu_l)|z|\right ]}^{f(W)},
    \label{Wf2}
\end{equation}
Using the Lambert W-function, we obtain 
\begin{equation}
I(z)=\frac{h\nu_l R_{\rm{rel}}}{\sigma(\nu_l)}\cdot W\left[\frac{\sigma(\nu_l)I_{\textrm{in}}}{h\nu_l R_{\rm{rel}}}\text{exp}\left(\frac{\sigma(\nu_l)I_\textrm{in}}{h\nu_l R_{\rm{rel}}}-n_{\text{Rb}}\sigma(\nu_l)|z|\right)\right].
\label{Iz}
\end{equation}
This result is in a convenient form because $W$ is a known special function, built in software packages such as Mathematica\textsuperscript{TM}. 
Due to the optical absorption, the light intensity decreases as a function of $z$ as shown in Fig.\,\ref{dv}(a).


In order to maximize the average polarization of spin source,
we should determine the optimal laser frequency.
Figure~\ref{dv}(b) shows the average polarization versus the laser frequency detuning. 
The average polarization reaches a maximum when the frequency detuning is zero.
Therefore, we adjust the laser frequency to the resonance frequency in our experiment.
Using a wavelength meter to monitor the frequency for a month,
the long-term frequency drift of the pump laser is measured to be
\begin{equation}
    \textrm{d}\nu_l\approx 0\pm0.45~\rm{GHz}.
\end{equation}

\begin{figure} [t] 
	\makeatletter
	\def\@captype{figure}
	\makeatother
	\includegraphics[scale=1.02]{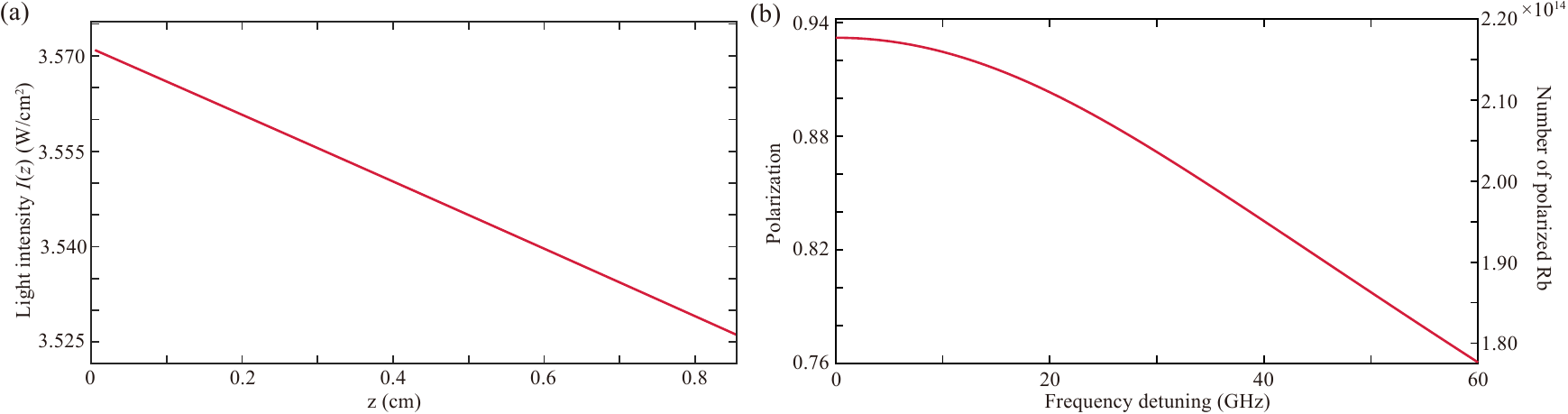}
	\caption{(a) Variation curve of the light intensity of the spin-source pump laser along $z$ [see Eq.~\eqref{Iz}]. (b) Variation curve of average polarization with detuning. The total number of polarized atoms reaches maximum when the frequency detuning is 0. The vertical axis on the left represents the variation curve of average polarization with the frequency detuning.}
	\label{dv}
\end{figure}
\end{itemize}

\begin{figure} [h] 
	\makeatletter
	\def\@captype{figure}
	\makeatother
	\includegraphics[scale=1]{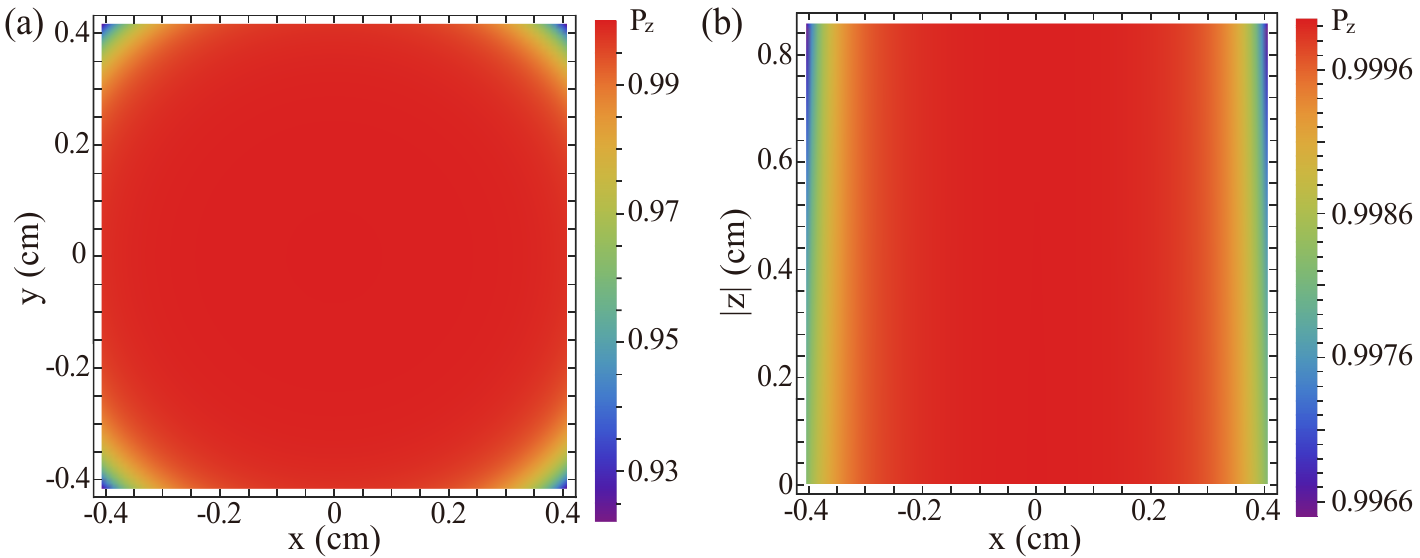}
	\caption{\textbf{Spatial distribution of spin polarization in the spin source cell by numerical simulation.} (a) Spatial polarization distribution in the $xy$ plane at $z=0$.  (b) Spatial polarization distribution in the $xz$ plane at $y=0$.}
	\label{Pz}
\end{figure}

Overall, with all the above parameters,
we obtain the spin polarization at each position in the spin-source cell.
The spatial polarization distribution of the spin source is shown in Fig.~\ref{Pz}.

\subsection{Systematic errors of spin-source parameters}

We first determine the mean value of the total number of polarized $^{87}$Rb spins using the mean values of all experimental parameters. 
Based on the spatial distribution of the spin polarization,
the total number of polarized $^{87}$Rb electron spins is obtained by integrating over the spin-source cell $N=\int_{V}\rho(\bm{r})\rm{d}\bm{r}$.
Subsequently, we obtain the error of the number of polarized $^{87}$Rb by determining and combining the systematic errors of all experimental parameters (see Table\,\ref{P}). 
Here, we take the laser power $p$ as an example to show how to derive the systematic error of the number of polarized $^{87}$Rb caused by the fluctuation of laser power.
The number density of polarized $^{87}$Rb $\rho(\bm{r})$ is re-estimated with the upper/lower limit of the pump power and mean values of other experimental parameters ( see Eq.\,\ref{rho} and Sec.\,\ref{SsB}).
Then, the number of polarized $^{87}$Rb is obtained by integrating over the spin-source cell 2.
The upper (lower) limit $N^+$ ($N^-$) on the number of polarized $^{87}$Rb is calculated by using the upper (lower) limit of the pump power $0.55\,\rm W$ ($0.45\,\rm W$),
i.e., $\Delta N^{\pm}=|N^{\pm}-N_0| \approx0.02\times10^{14}$,
where $N_0$ is the number of polarized $^{87}$Rb with the mean value of the pump power (0.5\,W).
The systematic errors caused by other parameters in Table \ref{P} are determined by using the same procedure.
The overall systematic error $\pm0.18\times10^{14}$ is derived by combining all the systematic errors in quadrature.
The total number of polarized $^{87}$Rb is $N\approx(2.18\pm0.18)\times10^{14}$.

\begin{table}[h]
\newcommand{\tabincell}[2]{\begin{tabular}{@{}#1@{}}#2\end{tabular}}
\begin{ruledtabular}
\caption {~~~Summary of experimental parameters and systematic error sources for the number of polarized Rb atoms.} 
\label{P}
\renewcommand{\arraystretch}{1.2}
\begin{tabular}{l c c}   
Parameter & Value & $\Delta N\left( \times 10^{14}\right)$ 
  \\
\hline
Length of spin source $L$ (mm) & $8.6\pm0.05$ & $\pm0.02$  \\
Width of spin source $W$ (mm) & $8.1\pm0.01$ & $\pm0.03$  \\ 
Height of spin source $H$ (mm) & $8.3\pm0.02$ & $\pm0.12$  \\
Number density of rubidium $n_{\text{Rb}}$ (cm$^{-3}$) & $(4.0\pm0.2)\times10^{14}$ & ${}^{+0.12}_{-0.11} $ \\
Full-width at half-maximum $\Gamma$ (GHz) & $6.88\pm 0.30$ & $<0.01$\\
Relaxation rate $R_{\text{rel}}$ ($s^{-1}$) & $524\pm18$ & $\pm0.01$\\
Beam waist $\omega$ ($\rm{mm}$) & $2.99\pm0.01$ & ${}^{+0.02}_{-0.03}$\\
Laser power $p$ (W) & $0.50\pm0.05$ & $\pm0.02$\\
Frequency detuning $\textrm{d}\nu_l$ (GHz) & $0\pm0.45$ & $<0.01$\\
[0.2cm]
Total Num of polarized rubidium $N$ $(\times 10^{14})$ & $2.18$
 & $\pm0.18$\\
\end{tabular} 
\end{ruledtabular}
\end{table}

\section{Numerical simulation of the exotic pseudomagnetic field}
\label{sec3}


Based on the number density of polarized $^{87}$Rb spins $\rho (\textbf{r})$ obtained in the previous section,
this section presents the numerical calculation of the pseudomagnetic field generated by the exotic interaction $V_{pp}$,
including the field strength and direction. 
The spin-dependent interaction studied here is~\cite{fadeev2019revisiting, dobrescu2006spin}
\begin{flalign}
      V_{pp}=-\frac{ g_p^e g_p^n}{16\pi m_{1}m_{2} }\left[(\hat{\bm{\sigma}}_1\cdot\hat{\bm{\sigma}}_2)\left(\frac{m_a}{ r^2}+\frac{1}{r^{3}}\right)-(\hat{\bm{\sigma}}_1\cdot\hat{\bm{r}})(\hat{\bm{\sigma}}_2\cdot\hat{\bm{r}})\left(\frac{m_a^2}{r}+\frac{3m_a}{r^2}+\frac{3}{r^3}\right)\right]e^{-m_a r},
\label{V3}
\end{flalign}
where $g_p^e g_p^n$ is the product of electron and neutron pseudoscalar coupling constants for $V_{pp}$, $\hat{\bm{\sigma}}_i$ is the spin vector of $i$th fermion and $m_i$ is its mass,
$r$ is the distance between the two interacting fermions,
$\hat{\bm{r}}$ is the corresponding unit vector,
and $m_a$ is the axion mass.

The exotic interaction $V_{pp}$ can induce energy shift of $^{129}$Xe spins in the spin sensor
 \begin{equation}
-\mu_{\rm{Xe}}\bm{\hat\sigma}_i\cdot\bm{B}_{a}^*=V_{pp},
\label{epm}
\end{equation}
where $\mu_{\text{Xe}}$ is the magnetic moment of $^{129}$Xe spin.
According to the Eqs.~(\ref{V3}) and (\ref{epm}),
the pseudomagnetic field $\bm{B}^*_{a}$ generated by a single electron spin is
\begin{flalign}
      \bm{B}^*_{a}=\frac{ g_p^e g_p^n}{16\pi\mu_{\text{Xe}}m_{1}m_{2} }\left[\hat{\bm{\sigma}}_2\left(\frac{m_a}{ r^2}+\frac{1}{r^{3}}\right)-\hat{\bm{r}}(\hat{\bm{\sigma}}_2\cdot\hat{\bm{r}})\left(\frac{m_a^2}{r}+\frac{3m_a}{r^2}+\frac{3}{r^3}\right)\right]e^{-m_a r}.
\label{b3}
\end{flalign}
With respect to all polarized electron spins and their spatial distribution in the spin source, the total pseudomagnetic field generated is
\begin{flalign}
     \bm{B}_{a}&=\int_{V}\rho(\bm{r}) \bm{B}^*_{a}(\bm{r})\rm{d}\bm{r}\notag \\
     &=\frac{ g_p^e g_p^n}{16\pi \mu_{\text{Xe}} m_{1}m_{2} }\int_{V}\rho(\bm{r})\left[\hat{\bm{\sigma}}_2\left(\frac{m_a}{ r^2}+\frac{1}{r^{3}}\right)-\hat{\bm{r}}(\hat{\bm{\sigma}}_2\cdot\hat{\bm{r}})\left(\frac{m_a^2}{r}+\frac{3m_a}{r^2}+\frac{3}{r^3}\right)\right]e^{-m_a r}\rm{d}\bm{r},
\label{B3}
\end{flalign}
where $V$ is the volume of spin source, $\rho(\bm{r})$ is the number density of polarized spins at the position $\bm{r}$ in the spin source.
According to the position of spin source and the direction polarized electron spins,
as shown in the Fig.~1 of the main text, $\bm{B}_{a}$ can be rewritten to
\begin{equation}
\bm{B}_{a}=B^y_{a}\hat{\bm{y}}+B^z_{a}\hat{\bm{z}}.
\end{equation}
Because the spin-based amplifier is insensitive to the pseudomagnetic field along $z$,
$B^z_{a}\hat{\bm{z}}$ is negligible and $B^y_{a}\hat{\bm{y}}$ is considered.

\begin{figure} [h] 
	\makeatletter
	\def\@captype{figure}
	\makeatother
	\includegraphics[scale=1.2]{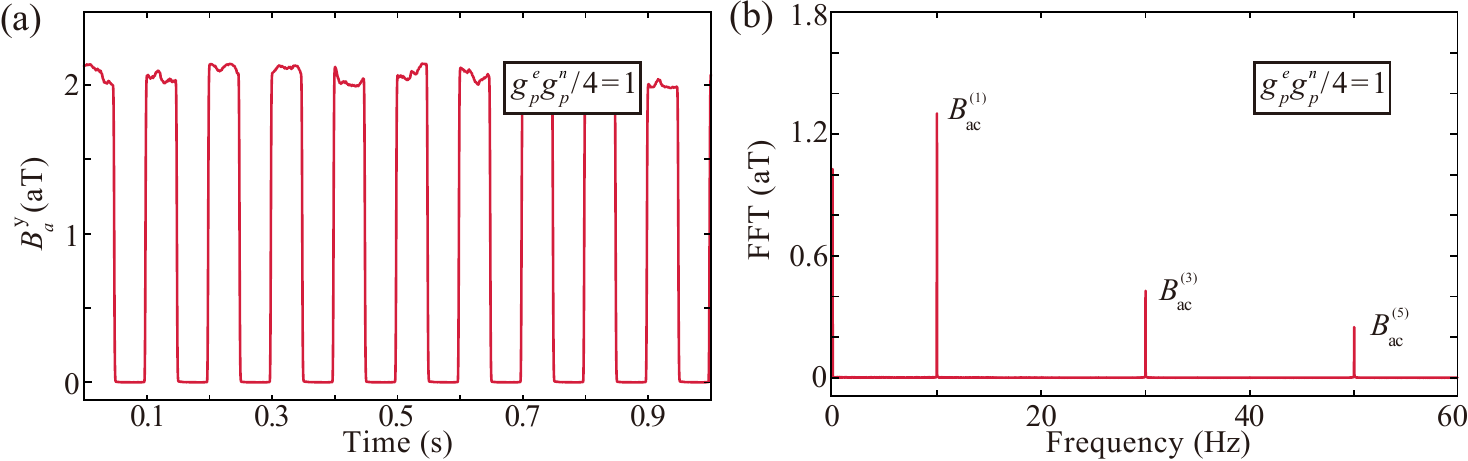}
	\caption{\textbf{Simulations of $V_{pp}$ produced by the spin source}. Assuming $g_p^e g_p^n/4=1$ and $m_a=0.1~$meV, the signal of the simulated pseudomagnetic field  $\bm{B}_{a}$ generated by the spin source is obtained. (a) Time-domain signal of $B^y_{a}$. (b) Fourier transformation spectrum of $B^y_{a}$. }
	\label{3_1}
\end{figure}

To generate the resonant pseudomagnetic field (an osillating field),
we use an optical chopper with 50$\%$ duty cycle to periodically block the pump light at the frequency of $\nu \approx 10.00$ Hz.
Because the $^{87}$Rb electron relaxation time ($\sim1$\,ms) is much shorter than the modulation period ($\sim100$\,ms), the modulated electron polarization can be thought of as changing instantaneously. 
Based on the variation of light intensity with time detected by PD1, we take $ g_p^e g_p^n/4=1$ and $m_a=0.1~$meV as an example to present the field strength and waveform of the pseudomagnetic field signal, as shown in Fig.~\ref{3_1}(a).
Furthermore, the frequency spectrum of the pseudomagnetic field signal can be obtained by performing discrete Fourier transform of the time domain signal.
As shown in Fig.~\ref{3_1}(b), the square wave signal $B^y_{a}$ contains trigonometric function signals with harmonic frequencies at $\nu$, 3$\nu$, 5$\nu$,... and the pseudomagnetic field along $y$ can be decomposed into
\begin{equation}
     B^y_{a}=\sum \limits_{N=\text{odd}} B^{(N)}_{\text{ac}} \cos(2 \pi N \nu t+\phi^{(N)}_0),
\label{Bacy}
\end{equation}
where $\phi^{(N)}_0$ and $B^{(N)}_{\text{ac}}$ is the phase and the strength of the $N$th harmonic field.
The ratios of the strengths are calculated to be $B_{\text{ac}}^{(1)}:B_{\text{ac}}^{(3)}:B_{\text{ac}}^{(5)}\approx13.0:4.3:2.5$.
In experiment, we can only detect one harmonic because of the narrow bandwidth $\approx$ 49\,mHz.
Therefore, we choose to detect the first harmonic $B_{\text{ac}}^{(1)}$ by tuning the resonant frequency $\nu_0$ of spin-based amplifier to 10.00\,Hz.

\section{Dipole magnetic field generated by the spin source}

In this section,
we theoretically calculate and experimentally measure the dipole magnetic field generated by the polarized $^{87}$Rb atoms in the spin source without magnetic shields.
If not properly shielded, the classical magnetic field is a spurious signal.
In the following, we discuss how to measure and remove the undesired field.

\subsection{Numerical calculation of the dipole magnetic field}

The dipole magnetic field generated by a single polarized $^{87}$Rb atom at position $\bm{r}$ can be expressed as
\begin{equation}
\bm{B}^*_d=\frac{\mu_0}{4\pi}\frac{3\hat{ \bm{r}}(\hat{\bm{r}}\cdot\bm{\mu}_e)-\bm{\mu}_e}{r^3}=\frac{\mu_0\mu_e}{4\pi}\frac{3\hat{\bm{r}}(\hat{\bm{r}}\cdot\hat{\bm{\sigma}}_2)-\hat{\bm{\sigma}}_2}{r^3},
\label{epm0}
\end{equation}
where $\mu_0$ is the vacuum permeability, $\hat{\bm{r}}$ is the unit vector of position vector $\bm{r}$, $\mu_e$ is the magnetic dipole moment of the electron spin, and $\hat{\bm{\sigma}}_2$ is its spin vector. 
Combining with the number density and direction of polarized $^{87}$Rb spins in the spin source, the strength and direction of the dipole magnetic field generated by the spin source can be calculated as
\begin{align}
\bm{B}_d &=\int_{V}\rho(\bm{r}) \bm{B}^*_d(\bm{r})\rm{d}\bm{r}\notag \\
&=\frac{\mu_0\mu_e}{4\pi}\int_{V}\rho(\bm{r})\left[\frac{3\hat{\bm{r}}(\hat{\bm{r}}\cdot\hat{\bm{\sigma}}_2)-\hat{\bm{\sigma}}_2}{r^3}\right]\rm{d}\bm{r},
\label{epm2}
\end{align}
where $V$ represents the integration space in the internal volume of the spin-source cell 2, $\rho(\bm{r})$ is the number density of polarized spins at the position $\bm{r}$. 
The dipole field at the center of the spin-based amplifier is $\bm{B}_d=B_d^x \hat{\bm{x}}+B_d^y \hat{\bm{y}}+B_d^z \hat{\bm{z}}$
with three components
\begin{equation}
\begin{aligned}
\label{dipoleM}
    B^x_d &=0.3~\textrm{pT},\\ 
    B^y_d & =-3.1~\textrm{pT},\\
   B^z_d & =2.8~\textrm{pT}.\\
\end{aligned}
\end{equation}



\subsection{Calibration of the dipole magnetic field}

\begin{figure} [h] 
	\makeatletter
	\def\@captype{figure}
	\makeatother
	\includegraphics[scale=1]{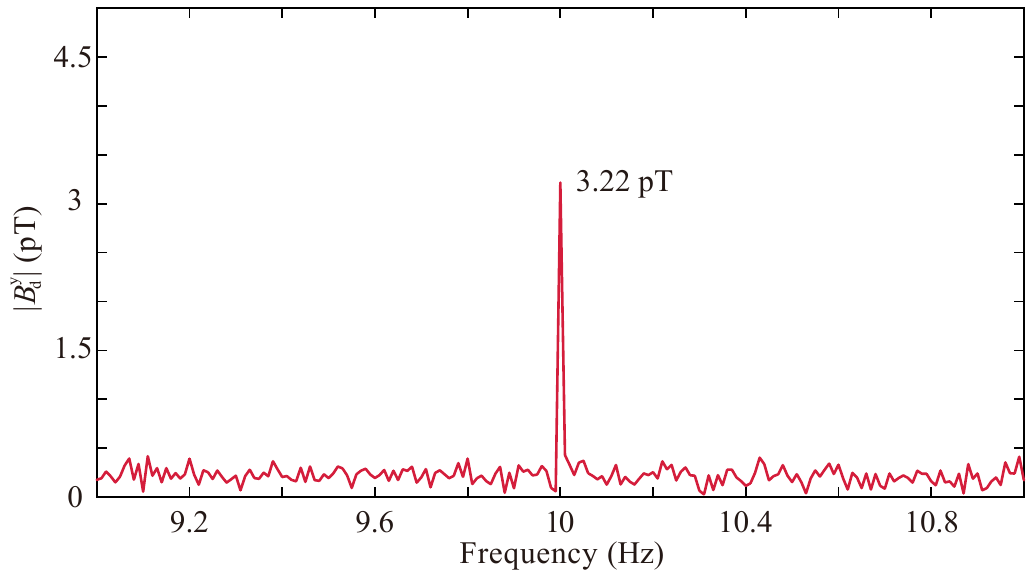}
	\caption{\textbf{Classical magnetic dipole field measured with a commercial miniature magnetometer without magnetic shields.} A frequency spectrum of the time-domain signal acquired with a QuSpin magnetometer. A dipole field $\approx 3.22$\,pT is measured}
	\label{Quspin}
\end{figure}

In order to accurately calibrate the strength of the dipole magnetic field,
we use a commercial miniature rubidium magnetometer (from QuSpin Inc.),
which is a centimeter-scale spin-exchange-relaxation-free (SERF) magnetometer.
This measurement is performed without small-size magnetic shields for the spin sensor and spin source (see Fig.\,\ref{Rsetup}).
The magnetic-field sensitivity is about 20\,fT/Hz$^{1/2}$ at frequencies ranging from 3 to 100\,Hz.
In order to measure the dipole field at the spin sensor,
we first remove the spin sensor and place the QuSpin magnetometer at the same position and then modulate the dipole field with an optical chopper that periodically blocks the high-power pump beam at a frequency $\nu \approx 10.00$\,Hz (see Sec.~\ref{Ss}).
As shown in Fig.\,\ref{Quspin},
there is a notable peak at about 10.00\,Hz and the dipole magnetic field is measured to be about 3.22\,pT,
in good agreement with the finite element analysis of magnetic field [see Eq.~(\ref{dipoleM})].

\subsection{Experimental elimination of the dipole magnetic field}
\label{shieldingD}

\begin{figure} [t] 
	\makeatletter
	\def\@captype{figure}
	\makeatother
	\includegraphics[scale=1.24]{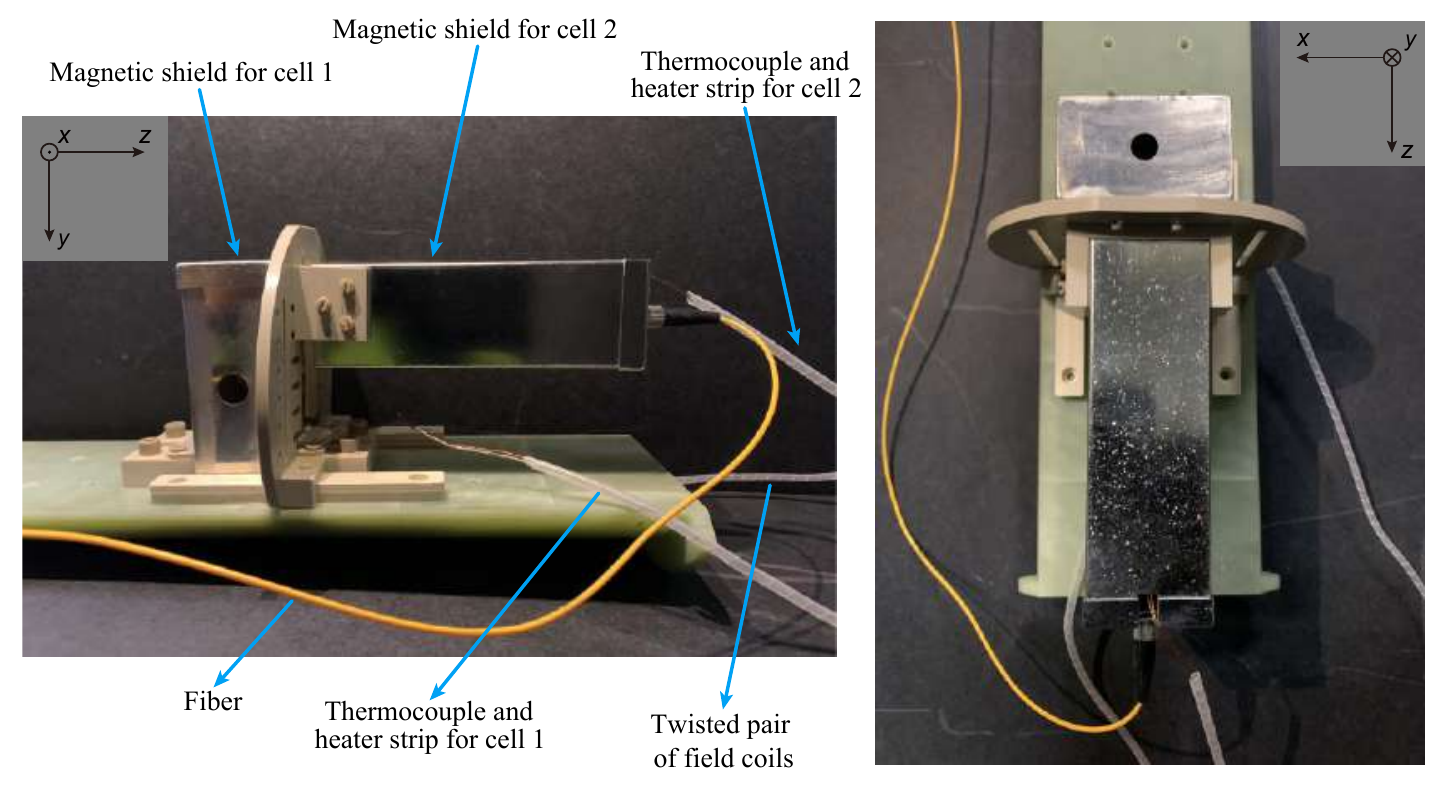}
	\caption{\textbf{Magnetic shields for eliminating the dipole magnetic field.}  Both of the spin sensor (cell 1) and the spin source (cell 2) are shielded with mu-metal magnetic shields.}
	\label{Rsetup}
\end{figure}

It is important to eliminate the dipole magnetic field that would generate the spurious signal hard to distinguish from the exotic interaction.
To do this,
we design and use small-size mu-metal shields for the spin sensor and spin source.
As shown in Fig.~\ref{Rsetup},
the spin sensor (cell 1) is magnetically shielded with a single layer of 1~mm thickness mu-metal shield,
inside which a set of Helmholtz coils is used to provide bias magnetic field.
The spin source (cell 2) is magnetically shielded with two layers of 0.5\,mm thickness mu-metal shields,
where the gap between the two layers is filled with aerogel for thermal insulation.

The shielding factor of mu-metal shields is calibrated with a fluxgate magnetometer.
The 1\,mm single-layer mu-metal shield for the spin sensor has a shielding factor of about $100$,
and the shielding factor of the two-layer shield for the spin source is about $4\times10^4$.
As a result, the total shielding factor is more than $10^6$.
The strength of the dipole magnetic field generated on the spin sensor should be less than $3\,\textrm{aT}$,
which is much less than the 24-hour magnetic-field detection limit of the spin sensor (76\,aT).
Therefore, the remaining magnetic dipole field can be neglected after adding the shields.

\section{Data analysis}

This section presents the procedure to determine the constraint on $g_p^e g_p^n$
and the data analysis of exotic-interaction-search experiments. 
The procedure includes five parts: the calibration experiment, the search experiment, analysis of mean values and statistical errors, analysis of systematic errors and, finally, the extraction of the constraint.

\begin{table}[h]
\newcommand{\tabincell}[2]{\begin{tabular}{@{}#1@{}}#2\end{tabular}}
\begin{ruledtabular}
\caption {~~~Summary of corrections to $|g_p^e g_p^n|/4$ (presented here for $m_a =0.1$\,meV) and systematic errors.} 
\label{tab:my_label_S2}
\renewcommand{\arraystretch}{1.2}
\begin{tabular}{l c c}   
Parameter & Value & $\Delta g_p^e g_p^n/4$ 
  \\
\hline
Position of cell 2 $x$ (mm) & $1.9\pm0.2$ & ${}^{<0.01}_{-0.04}$  \\[0.15cm]
Position of cell 2 $y$ (mm) & $-23.1\pm1.0$ & ${}^{-0.98}_{+1.23}$  \\ [0.15cm]
Position of cell 2 $z$ (mm) & $41.9\pm0.3$ & ${}^{+0.75}_{-0.69}$  \\ [0.15cm]
Num. of polarized Rb ($10^{14}$) & $2.18\pm0.18$ & ${}^{-0.41}_{+0.44}$ \\[0.15cm]
Phase delay $\phi$ (deg) & $102.6\pm0.6$ & ${}^{-1.45}_{+1.45}$\\[0.15cm]
Calib. const. $\alpha$ (V/nT) & $2.11^{+0.002}_{-0.386}$ & ${}^{-0.01}_{+1.20}$\\[0.25cm]

Final $|g_p^e g_p^n|/4$ & $5.3$
 & $\pm 48.5 \ (\text{statistical}) $\\

$(m_a =0.1~\text{meV})$ &  & $\pm 2.4 \ (\text{systematic})
$
\end{tabular} 
\end{ruledtabular}
\end{table}

\subsection{Calibration experiment}
\label{secC1}


To analyze the strength and phase of the pseudomagnetic field signal from $V_{pp}$,
we calibrate experimental parameters in advance,
including the position of the spin source $\bm{r}$, number density of polarized spins in the spin source $\rho(\bm{r})$,
phase delay $\phi$ and calibration constant $\alpha$ of the spin-based amplifier, etc.
With the position $\bm{r}$ and the number density of polarized spins $\rho(\bm{r})$,
the magnitude of the pseudomagnetic field generated by the spin source can be calculated according to Eq.~(\ref{B3}). 
In addition,
the phase delay $\phi$ and calibration constant $\alpha$ are used to determine the amplitude and phase of the output-voltage signal generated by the pseudomagnetic field.
Table \ref{tab:my_label_S2} shows the mean values and uncertainties of the physical parameters above.
In the following, we explain the details of such parameters.

\begin{itemize}
\item[(1)] \textbf{Calibration of the position of spin source}.
We measure the position of the spin source $\bm{r}=(x, y, z)$.
The coordinate axis directions are shown in Fig.~\ref{setup}.
The position of the center of the spin source (cell 2) relative to the origin of coordinate (the center point of the cell 1) is
\begin{equation}
\begin{aligned}
    x &\approx 1.9\pm0.2~\textrm{mm},\\ 
    y &\approx -23.1\pm1.0~\textrm{mm},\\
    z &\approx 41.9\pm0.3~\textrm{mm}.\\
\end{aligned}
\end{equation}
The length, width and height of the spin source are shown in Table \ref{P}.
According to the center position $\bm{r}$ and size of the spin source,
the position of a single micro-element relative to the origin is determined and further used to determine the number density of polarized $^{87}$Rb atoms for this micro-element.
\end{itemize}

\begin{itemize}
\item[(2)] \textbf{Calibration of number density of polarized spins in the spin source}.
In order to calculate the amplitude of the pseudomagnetic field generated by the spin source,
we need to calibrate the number density of polarized $^{87}$Rb atoms at each position in the spin source $\rho(\bm{r})=n_{\text{Rb}}\cdot P_z(\bm{r})$ [see Eq.~(\ref{B3})].
The calibration procedure is described in Sec.~\ref{SsB} and we only present the result here. 
The total number of polarized $^{87}$Rb atoms in the spin source is calculated based on $N =\int_{V}\rho(\bm{r})\text{d}\bm{r}$ and presented in Table \ref{P}
\begin{equation}
     N \approx (2.18\pm0.18)\times 10^{14}.
\end{equation}

\end{itemize}

\begin{itemize}
\item[(3)] \textbf{Determination of the calibration constant $\alpha$}.
The calibration constant $\alpha$ ($\textrm{V/nT}$) of the spin-based amplifier describes the conversion coefficient between the strength of input magnetic field and the output voltage signal of the spin-based amplifier.
To measure $\alpha$,
we apply a known oscillating magnetic field along $y$ and measure the amplitude of the output-voltage signal.
The calibration constant $\alpha$ is determined as
\begin{equation}
    \alpha \approx 2.11^{+0.002}_{-0.386}~\textrm{V/nT}.
    \label{deltaa}
\end{equation}
The uncertainty of the calibration constant $\alpha$ is determined by two aspects:

\item The intrinsic instability of the spin-based amplifier $\delta \alpha_{\textrm{int}}$.
The intrinsic instability $\delta \alpha_{\textrm{int}}$ can be caused by technical sources of random fluctuations, such as the bias magnetic field along $z$,
variations in vapor-cell temperature and optical power, etc.
To calibrate $\delta \alpha_{\textrm{int}}$,
we measure the output signal of the spin-based amplifier for one hour with an oscillating magnetic field applied.
According to the fluctuation of the output signal strength,
we obtain the intrinsic instability $\delta \alpha_{\textrm{int}}\approx \pm0.002~\textrm{V/nT}$.

\item The external instability of input-signal frequency $\delta \alpha_{\textrm{ext}}$ from the instability of the chopper.
The instability of pseudomagnetic field frequency $\delta\nu$ is caused by the fluctuation of chopper frequency, which is calibrated as $\delta\nu\approx \pm10 ~\textrm{mHz}$ by monitoring the frequency stability of output chopper signal at PD2 for a long time.
The mismatch between the input-signal frequency and the resonance frequency of the spin-based amplifier would weaken the amplification effect.
The calibration constant $\alpha(\nu)$ can be described by the lineshape formula
\begin{equation}
    \alpha(\nu) =\alpha(\nu_0) \dfrac{\Lambda/2}{\sqrt{(\nu-\nu_0)^2+(\Lambda/2)^2}},
    \label{line_shape}
\end{equation}
where $\Lambda\approx 28 ~\textrm{mHz}$.
Assuming $(\nu-\nu_0)\approx\delta\nu$ in Eq.~(\ref{line_shape}), the external uncertainty can be $\delta\alpha_{\textrm{ext}}\approx -0.386~\textrm{V/nT}$.
Combining $\delta \alpha_{\textrm{int}}$ and $\delta \alpha_{\textrm{ext}}$, the total uncertainty $\delta \alpha$ of the spin-based amplifier is determined as $\alpha \approx 2.11^{+0.002}_{-0.386}~\textrm{V/nT}$.


\end{itemize}

\begin{itemize}
\item[(4)] \textbf{Calibration of the phase delay $\phi$}. The phase delay of the spin-based amplifier $\phi$ denotes the phase difference between the input-magnetic-field signal and the output-voltage signal of the spin-based amplifier.
In the actual measurement process,
we apply an oscillating magnetic field along $y$ with known phase $\phi_0$ and frequency $\nu$.
Then, the phase delay of the spin-based amplifier is obtained by measuring the difference between the phase of the output voltage signal and the phase of the input magnetic field.
The phase delay is determined as
\begin{equation}
    \phi \approx 102.6\pm 0.6~\textrm{deg},
\end{equation} 
where the uncertainty of phase delay $\phi$ is obtained by repeating the above process and calculating the corresponding variance.

\end{itemize}

The experimental parameters determined in the calibration experiments are listed in the second column of Table~\ref{tab:my_label_S2}. 
Using the calibration constant $\alpha$ and the phase delay $\phi$ of the spin-based amplifier,
the output voltage signal of the spin-based amplifier can be determined as $\alpha B^y_{ac}\cos(2\pi\nu t+\phi_0-\phi)$.
Then, the pseudomagnetic field can be obtained from the experimental data as discussed in Sec.\,\ref{secC2} and \ref{secC3} .

\subsection{Search experiment}
\label{secC12}

Throughout the experiment, the spin-based amplifier is tuned to match the optical chopping frequency of the spin-source pump laser, i.e., $\nu_0 \approx \nu \approx 10.00$\,Hz.
Due to the narrow bandwidth (49\,mHz) of the amplifier, only the first harmonic of the pseudomagnetic field can be amplified and the effect of other harmonics is negligible (see Sec.~\ref{sec3}).
The data were collected in six-hour batches,
after which the spin-based amplifier was readjusted to optimize the magnetic-field sensitivity by tweaking the bias field, etc.
(see Sec.\,\ref{Ssen}).
While recording search data,
the parameters of the spin source were monitored, such as the chopper frequency and pump power.
The total duration of the search experiment was about 24\,h.

\subsection{Analysis of mean values and statistical errors}
\label{secC2}

We now describe how to obtain the constraints on the product of coupling constants $|g_p^e g_p^n|/4$ from the experimental search data.
A ``lock-in'' detection scheme is used to extract the weak signal from the pseudomagnetic fields with a known carrier frequency from noisy signal~\cite{su2021search,ji2018new}.
The procedure is divided into two steps.

\begin{itemize}
\item[(1)] \textbf{Step~1: Obtain the product of coupling constants $g_p^e g_p^n/4$ for one-hour batches}.

\item After applying a band-pass filter around the resonance frequency on the data,
the noise such as power-frequency noise (50~Hz, 100~Hz, ...) can be filtered out.
The filtered time-domain data are further separated into segments of one period $S(t)^i$.

\item The normalized reference signal $\cos{(2 \pi \nu t+\phi_0^{(1)}-\phi)}$ of every individual period is simulated based on Eq.~(\ref{Bacy}),
corresponding to $g_p^e g_p^n/4=1$ for the mass $m_a=0.1$~meV based on experimental parameters in Table~\ref{tab:my_label_S2}.

\item Using ``lock-in'' detection scheme,
we extract every product of coupling constants $g_p^e g_p^n/4$ for every period $S(t)^i$ ($T \approx 100$~ms)
\begin{equation}
\label{S30}
    (g_p^e g_p^n)^{i}/4= \frac{1}{\alpha B^{(1)}_{\textrm{ac}}} \frac{\int^{T}_0 \cos{(2 \pi \nu t+\phi_0^{(1)}-\phi)} S(t)^i \textrm{d} t}{\int^{T}_0 \cos^2{(2 \pi \nu t+\phi_0^{(1)}-\phi)} \textrm{d} t}, 
\end{equation}
where $B^{(1)}_{\textrm{ac}}$ is the first harmonic corresponding to $g_p^e g_p^n/4=1$ (see Sec.~\ref{sec3}).
Figure~\ref{gauss} shows the graph of all products of coupling constants $g_p^e g_p^n/4$ obtained for the-first-hour batches.
The fit with Gaussian distribution to the histogram gives the mean value and the standard error of the products of coupling constants $g_p^e g_p^n/4$,
\begin{equation}
    (g_p^e g_p^n)^{\textrm{1~h}}/4 \approx 52.7 \pm 218.0_{\textrm{stat}},
\end{equation}
where the superscript 1~h represents first one-hour batch.
By repeating the above three steps,
24 products of coupling constants $g_p^e g_p^n/4$ are obtained.

\begin{figure} [t] 
	\makeatletter
	\def\@captype{figure}
	\makeatother
	\includegraphics[scale=0.85]{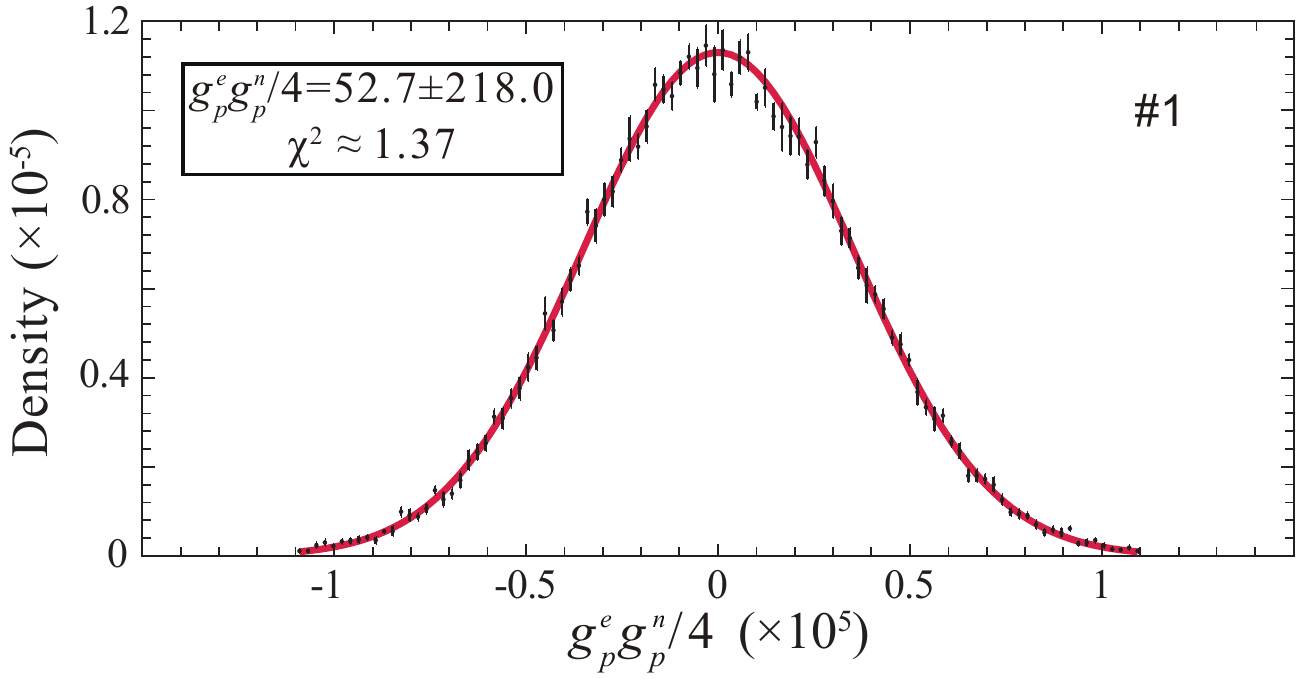}
	\caption{\textbf{The graph of the potential experimental coupling strength} $(g_p^e g_p^n)^{\text{1~h}}/4$. Distribution of the experimental coupling strength $(g_p^e g_p^n)^{\textrm{1~h}}/4$ of the-first-hour data. The red solid line is a fit to a Gaussian distribution. The average and the standard error of the coupling strength is $(g_p^e g_p^n)^{\textrm{1~h}}/4 \approx 52.7\pm218.0$. The $\chi^2 \approx 1.37$ represents a valid fitting.}
	\label{gauss}
\end{figure}

\item[(2)]
\textbf{Step~2: We use the reduced $\chi^2$ statistic to obtain the mean value and standard error of the products of coupling constants $(g_p^e g_p^n)/4$ for 24-hour batches}.
The weighted average and the standard error of the product of coupling constants $|g_p^e g_p^n|/4$ is defined as~\cite{lee2018improved}
\begin{equation}
\begin{aligned}
    &\left<x\right>=\left|\frac{\Sigma^N_{i=1}\varpi^i x^{i}}{\Sigma^N_{i=1}\varpi^i}\right|,\\
    &\Delta x=\frac{1}{\sqrt{\Sigma^N_{i=1}\varpi^i}},
\end{aligned}
\end{equation}
where $x$ represents $|g_p^e g_p^n|/4$, $x^{i}\pm\sigma^i$ is the mean value and the standard error of coupling constants $|g_p^e g_p^n|^i/4$ for one hour, the weights are defined as $\varpi^i=\left(1/\sigma^i\right)^2$.
The values with the smaller $\sigma^i$ carry the larger statistical weight.
The weighted reduced $\chi^2$ is
\begin{equation}
    \chi^2=\frac{\Sigma^N_{i=1}\varpi^i}{(\Sigma^N_{i=1}\varpi^i)^2-\Sigma^N_{i=1} (\varpi^i)^2} \cdot \sum_{i=1}^N \frac{\varpi^i(x^i-\left<x\right>)^2}{(\sigma^{i})^2}.
\end{equation}
Based on the $\chi^2$ statistic,
the mean value and standard error of the products of coupling constants $|g_p^e g_p^n|/4$ for 24-hour batches is $5.3\pm48.5_{\rm{stat}}$ and the weighted reduced $\chi^2$ is $1.65$.

\end{itemize}

\subsection{Analysis of systematic errors}
\label{secC3}

As was shown above,
the mean value and the standard error of the parameters are obtained.
Then we can determine the mean value and statistical errors of the coupling strength (the product of coupling constants $|g_p^e g_p^n|/4$). 
This section analyzes the error of the coupling strength caused by the systematic errors of experimental parameters (see Table~\ref{tab:my_label_S2}). 

\begin{figure} [h] 
	\makeatletter
	\def\@captype{figure}
	\makeatother
	\includegraphics[scale=1.0]{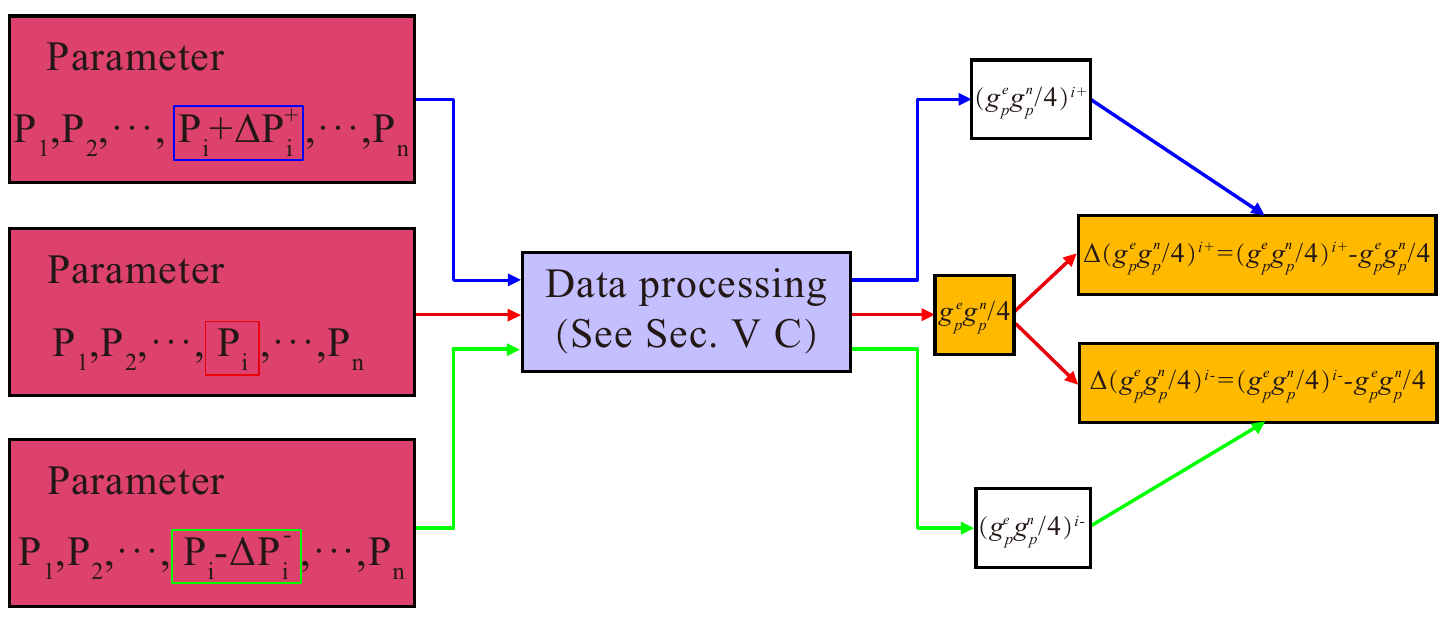}
	\caption{\textbf{Procedures for determining systematic errors.} The left part represents the average value and standard error of parameters. The middle part represents the data processing (see Sec.~\ref{secC2}). The right part represents the determined systematic errors.}
	\label{Procedure2}
\end{figure}

Consider the example of calibration constant $\alpha$,
we determine its systematic error using three steps shown in Fig.~\ref{Procedure2}:

\begin{itemize}
\item[(1)] Obtaining the reference signal. By substituting the upper limit of the calibration constant $\alpha=2.110+0.002~\textrm{V/nT}$ and the mean values of other parameters (see Table~\ref{tab:my_label_S2}) into Eq.\,(\ref{S30}),
we obtain the reference signal corresponding to the upper limit of the calibration constant. 

\item[(2)] Extracting the coupling strength. By using the reference signal obtained above,
we can extract the corresponding coupling strength $(|g_p^e g_p^n|/4)^{i+}$ by employing the ``lock-in" detection scheme. 

\item[(3)] Determine the systematic error. The systematic error caused by the upper limit of the calibration constant is determined by the difference $\Delta (|g_p^e g_p^n|/4)^{i+}=(|g_p^e g_p^n|/4)^{i+}-|g_p^e g_p^n|/4\approx-0.01.$
Similarly, we can obtain the systematic error corresponding to the lower limit of the calibration constant $\alpha=2.110-0.386~\textrm{V/nT}$,
which is given by  $\Delta (|g_p^e g_p^n|/4)^{i-}=(|g_p^e g_p^n|/4)^{i-}-|g_p^e g_p^n|/4\approx+1.20$.

\end{itemize}

Based on these procedures, the systematic errors caused by other calibration parameters in Table~\ref{tab:my_label_S2} are determined.
Finally, we can derive the overall systematic error $\pm 2.4_{\textrm{sys}}$ by combining all the systematic errors in quadrature, and the final coupling strength is
\begin{equation}
|g_p^e g_p^n|/4\approx5.3\pm48.5_{\textrm{stat}}\pm 2.4_{\textrm{sys}}.
\end{equation}. 

\subsection{Extraction of the constraint}

Based on the above data analysis,
the constraint on coupling strength at a specific mass $m_a$ is determined.
Taking $m_a=0.1~\textrm{meV}$ as an example, the constraints on $|g_p^e g_p^n|/4$ at confidence level of 95$\%$ (corresponding to 1.96$\sigma$) can be determined as
\begin{equation}
\label{}
5.3+1.96\times\sqrt{48.5^2_{\textrm{stat}}+2.4^2_{\textrm{sys}}}\approx100.
\end{equation}
Further, by repeating the data analysis in Secs.~\ref{secC2} and \ref{secC3} and changing mass $m_a$,
the constraints for the entire explored mass range are obtained,
as shown in Fig.~4 in the main text. 
Finally, we establish the constraints on the exotic spin-spin interaction between polarized electron and neutron spins in the axion window (1$~\mu$eV-1$~$meV),
which corresponds to a force range from 0.2$~$mm to 20$~$cm.
In particular, we obtain the most stringent constraints on $|g_p^e g_p^n|/4$ for the mass range from 0.03$~$meV to 1$~$meV.

\bibliographystyle{naturemag}
\bibliography{SIrefs}